%% file: main.tex
\documentclass[journal]{IEEEtran}
\usepackage[colorlinks, citecolor=blue]{hyperref}

\usepackage{cite}
\usepackage[table]{xcolor}
\usepackage{url,hyperref,booktabs}

\usepackage{pifont}
\usepackage{makecell}

\ifCLASSINFOpdf
  \usepackage[pdftex]{graphicx}
  \DeclareGraphicsExtensions{.pdf,.jpeg,.png}
  \usepackage{epstopdf}
  \usepackage{subcaption}
\else
  \usepackage[dvips]{graphicx}
\fi
\usepackage{amsmath}
\usepackage{amssymb}

\usepackage{multirow}
\usepackage[switch]{lineno}
\usepackage{array}

\usepackage{stfloats}
\usepackage[symbol]{footmisc}

\makeatletter
\def\endthebibliography{%
  \def\@noitemerr{\@latex@warning{Empty `thebibliography' environment}}%
  \endlist
}
\makeatother

\begin{document}
% \linenumbers
% \title{A Time-Frequency Attention Module for Neural Speech Enhancement}
% \title{Ada-FE: A Bio-Inspired Adaptive Front-End With Dynamic Neural Control for Audio Processing}
\title{\textcolor{black}{Should Audio Front-ends be Adaptive? \textcolor{black}{Comparing} Learnable and Adaptive Front-ends}}
%Ada-FE: An Adaptive Front-End With Neural Feedback Control for Audio Processing}
% Programmatic Grant No. A18A2b0046 from the Singapore Government’s Research, Innovation and Enterprise 2020 plan (Advanced Manufacturing and Engineering domain)
\author{Qiquan~Zhang,~\IEEEmembership{Member,~IEEE,}
        Buddhi Wickramasinghe, %~\IEEEmembership{Member,~IEEE,}
        Eliathamby Ambikairajah,~\IEEEmembership{Life Senior Member,~IEEE,}
        Vidhyasaharan Sethu,~\IEEEmembership{Member,~IEEE,}
        % Eliathamby Ambikairajah,~\IEEEmembership{Senior Member,~IEEE,} 
        and Haizhou Li,~\IEEEmembership{Fellow,~IEEE}
\thanks{
Manuscript received date; revised date. 
% This work was supported in part by ARC Discovery Grant DP1900102479, in part by the Research Foundation of Guangdong Province under Grant 2019A050505001, in part by the Internal Project of Shenzhen Research Institute of Big Data under Grant T00120220002, and in part by the Guangdong Provincial Key Laboratory of Big Data Computing under the Grant B10120210117-KP02. 
% in part by University Development Fund under Grants UDF01002333 and UF02002333, {in part by the Research Foundation of Guangdong Province under Grant 2019A050505001}, in part by the Science and Engineering Research Council, Agency for Science, Technology and Research (A*STAR), Singapore, through the National Robotics Program under Human-Robot Interaction Phase 1 under Grant 1922500054, and in part by the Deutsche Forschungsgemeinschaft (DFG, German Research Foundation) under Germany’s Excellence Strategy (University Allowance, EXC 2077, University of Bremen). \textit{(Corresponding Author: Xinyuan Qian)}.
}
\thanks{Qiquan Zhang, Eliathamby Ambikairaiah, and Vidhyasaharan Sethu are with the School of Electrical Engineering and Telecommunications, The University of New South Wales, Sydney, 2052, Australia~(e-mail: {qiquan.zhang}@unsw.edu.au; {e.ambikairajah}@unsw.edu.au).}
\thanks{Buddhi Wickramasinghe is with the School of Electrical and Computer Engineering, Purdue University, West Lafayette IN, USA~(e-mail: {wwickram}@purdue.edu).}
% \thanks{Qiquan Zhang is also with the Department of Electrical and Computer Engineering, National University of Singapore, 119077, Singapore.}
% \thanks{Xinyuan Qian is with the School of Computer and Communication Engineering, University of Science and Technology Beijing, Beijing 100083, China, and also with The Chinese University of Hong Kong, Shenzhen 518172, China (e-mail: qianxy@ustb.edu.cn)}
% \thanks{Zhaoheng Ni is with Meta AI, New York, 10003, United States of America (e-mail: {zni}@fb.com).}
% \thanks{Aaron Nicolson is with Australian e-Health Research Centre, CSIRO, Australia (e-mail: {aaron.nicolson}@csiro.au).}
% \thanks{Eliathamby Ambikairaiah and Vidhyasaharan Sethu are with the School of Electrical Engineering and Telecommunications at the University of New South Wales, Sydney, 2052, Australia (e-mail: {e.ambikairajah}@unsw.edu.au).}
\thanks{Haizhou Li is with the Guangdong Provincial Key Laboratory of Big Data Computing, The Chinese University of Hong Kong (Shenzhen), 518172 China, and also with Shenzhen Research Institute of Big data, Shenzhen, 51872 China (e-mail: {haizhouli}@cuhk.edu.cn). 
% also with the Department of Electrical and Computer Engineering, National University of Singapore, 119077, Singapore; also with the University of Bremen, 28359 Germany and also with Kriston AI Lab, Xiamen, China (e-mail: {haizhouli}@cuhk.edu.cn).
}
% \thanks
}% <-this % stops a space
% \thanks{Manuscript received April 19, 2005; revised August 26, 2015.}}
% \cortext[cor1]{Corresponding author}


\markboth{Journal of \LaTeX\ Class Files,~Vol.~14, No.~8, August~2019}%
{Shell \MakeLowercase{\textit{et al.}}: Bare Demo of IEEEtran.cls for IEEE Journals}



\maketitle

\begin{abstract}
% Speech enhancement plays an essential role in a wide range of speech processing applications. Recent studies on speech enhancement tend to investigate how to effectively capture the long-term contextual dependencies of speech signals to boost performance. However, these studies generally neglect the time-frequency (T-F) distribution information of speech spectral components, which is equally important for speech enhancement. In this paper, we propose a simple yet very effective network module, which we term the T-F attention \textcolor{black}{(TFA) module}, that uses two parallel attention branches, i.e., time-frame attention and frequency-channel attention, to explicitly exploit position information to generate a 2-D attention map to characterise the salient T-F speech distribution. We validate our \textcolor{black}{TFA module} as part of two widely used backbone networks (residual temporal convolution network and Transformer) and conduct speech enhancement with four most popular training objectives. Our extensive experiments demonstrate that our proposed \textcolor{black}{TFA module} consistently leads to substantial enhancement performance improvements in terms of the five most widely used objective metrics, with negligible parameter overheads. In addition, we further evaluate the efficacy of speech enhancement as a front-end for a downstream speech recognition task. Our evaluation results show that the \textcolor{black}{TFA module} significantly improves the robustness of the system to noisy conditions. 

\textcolor{black}{Hand-crafted features, such as Mel-filterbanks, have traditionally been the choice for many audio processing applications. Recently, there has been a growing interest in learnable front-ends that extract representations directly from the raw audio waveform. \textcolor{black}{However, both hand-crafted filterbanks and current learnable front-ends lead to fixed computation graphs at inference time, failing to dynamically adapt to varying acoustic environments, a key feature of human auditory systems.} To this end, we explore the question of whether audio front-ends should be adaptive by comparing the Ada-FE front-end (a recently developed adaptive front-end that employs a neural adaptive feedback controller to dynamically adjust the Q-factors of its spectral decomposition filters) to established learnable front-ends. Specifically, we systematically investigate learnable front-ends and Ada-FE across two commonly used back-end backbones and a wide range of audio benchmarks including speech, sound event, and music. The comprehensive results show that our Ada-FE outperforms advanced learnable front-ends, and more importantly, it exhibits impressive stability or robustness on test samples over various training epochs.
}

\end{abstract}


% \fi
\IEEEpeerreviewmaketitle

\section{Introduction}\label{sec:1}

\IEEEPARstart{T}{he} \textcolor{black}{last decade has witnessed substantial success in deep learning to speech and audio processing. More recently, audio representation learning has received increasing interest, which seeks to extract discriminative and robust audio features for downstream tasks. It has demonstrated impressive advances in a variety of audio tasks, such as phone recognition~\cite{zeghidour2018learning}, automatic speech recognition (ASR)~\cite{asr2015}, speaker recognition~\cite{sincnet}, spoofing speech detection~\cite{dinkel2017end}, sound event classification~\cite{esc50}, music understanding~\cite{wang2022towards}, and speech separation~\cite{pariente2020filterbank}.}

% Feature extraction is one critical step for building a deep learning-based audio processing system.

\begin{figure}[!t]
% \vspace{-1.3em}
\centering
\begin{subfigure}[t]{0.97\columnwidth}
\centerline{\includegraphics[width=0.99\columnwidth]{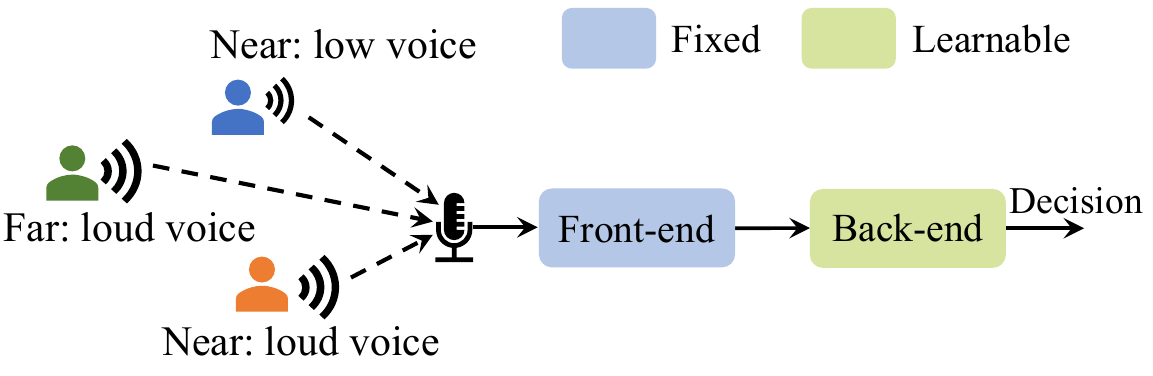}}
\caption{\textcolor{black}{Fixed during training and inference.}}
\hfill
\label{fig1:1}
\end{subfigure}
\begin{subfigure}[t]{0.98\columnwidth}
\centerline{\includegraphics[width=0.99\columnwidth]{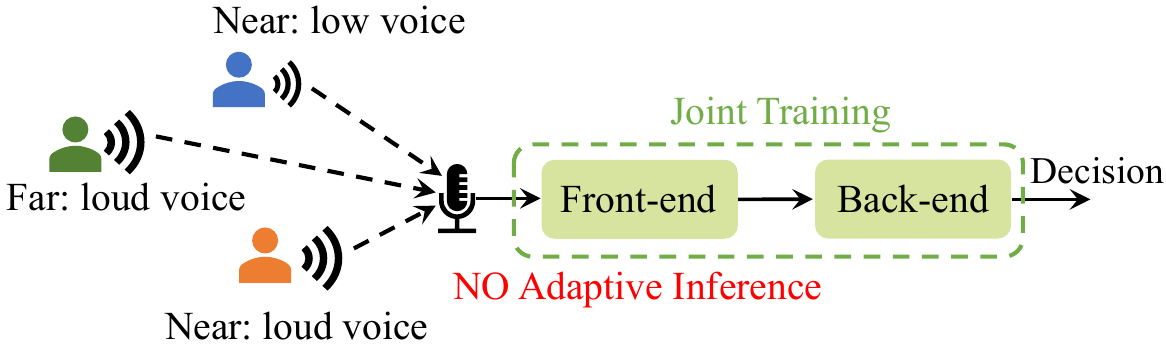}}
\caption{\textcolor{black}{Learnable during training, fixed at inference time}}
\hfill
\label{fig1:2}
\end{subfigure}
\begin{subfigure}[t]{0.98\columnwidth}
\centerline{\includegraphics[width=0.99\columnwidth]{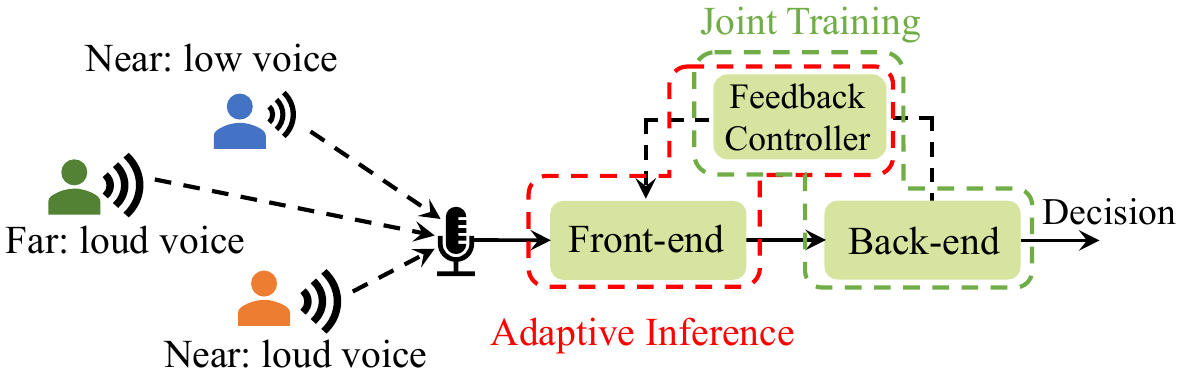}}
\caption{\textcolor{black}{Learnable during training, adaptive at inference time}}
\label{fig1:3}
\end{subfigure}

\caption{\textcolor{black}{
\textcolor{black}{\textcolor{black}{Illustration of (a) fixed, (b) learnable but non-adaptive, and (c) learnable and adaptive audio front-ends.} In real-world acoustic scenarios, speech and audio signals are inevitably shaped by varying acoustic conditions during transmission. The input speech level, for instance, varies with the distance between the speakers and the microphone and the voice levels of the speakers. (a) Fixed (hand-crafted) front-ends extract feature using fixed filters during training and inference regardless of the acoustic scene. (b) Existing neural front-ends learn a common model to deal with the different acoustic scenes present in the training data but cannot change to compensate for previously unseen acoustic scenarios at test time}. (c) Our adaptive front-end employs a neural controller to dynamically adjust the filter response (or shape) to varying acoustic conditions during training and inference.
% Illustration of neural audio front-ends. In real-world acoustic scenarios, speech and audio signals are inevitably shaped by varying acoustic conditions during transmission. The input speech level, for instance, varies with the distance (related to the transfer function) between the speakers and the microphone and the voice levels of the speakers. (a) Existing learnable front-ends are fixed at inference time, regardless of the input speech level. (b) Our adaptive front-end employs a feedback controller to dynamically adjust the filter response (or shape) to varying acoustic conditions during training and inference.
} 
% \textcolor{red}{Redraw this figure to emphasize noisy environment instead of low/loud voice}
}
\vspace{-0.8em}
\label{fig1}
\end{figure}

% The hand-crafted Mel-filterbank features have established remarkable success in multiple application domains and remained the features of choice for many state-of-the-art audio processing systems~\cite{conformer}. Mel-filterbanks first apply the short-term Fourier transform (STFT) to the raw waveform, followed by a squared modulus operation. The resulting STFT power spectrogram is then passed through a triangular band-pass filterbank spaced according to the Mel-scale to model the non-linear human perception, which provides robustness and shifts-invariance to deformations~\cite{anden2014deep}. Finally, a logarithmic compression function is typically employed to reduce the dynamic range of filterbank energy to match the human response to loudness.

\textcolor{black}{Until fairly recently audio front-ends have been fixed (Figure 1a), with the Mel filterbank being the front-end of choice} across numerous audio processing systems~\cite{conformer}. Mel-filterbanks first apply the short-term Fourier transform (STFT) to the raw waveform, followed by a squared modulus operation. The resulting STFT power spectrogram is then passed through a triangular band-pass filterbank spaced according to the Mel-scale to model the non-linear human perception, which provides robustness and shifts-invariance to deformations~\cite{anden2014deep}. Finally, a logarithmic compression function is employed to reduce the dynamic range of filterbank energy to match the human response to loudness. \textcolor{black}{However, Mel-filterbanks suffer from limitations of fixed representations, such as a lack of robustness to wide variations in loudness.} Similarly, Mel-filterbanks are incapable of providing robust pitch harmonics to children's ASR systems~\cite{pitch}. To address these limitations, there has been an increased popularity of employing neural filter layers to learn a front-end directly from raw waveforms to replace hand-crafted filterbanks~\cite{he2016deep,luo2019conv,demucs}. Such a way allows for the optimization of neural filterbanks specifically for the audio tasks at hand in an end-to-end fashion.

\textcolor{black}{In recent years, multiple types of neural filter layers have been investigated to build learnable front-ends, involving standard convolutional layers~\cite{conv15,7178847}, dilated convolutional layers~\cite{schneider2019wav2vec}, and parametric neural filters such as Gaussian~\cite{gauss}, Gammatone~\cite{sainath15_interspeech}, Sinc~\cite{sincnet,pariente2020filterbank}, and Gabor filters~\cite{leaf}.} Parametric neural filters, in comparison to standard convolutional layers that learn all the filter parameters, only involve several trainable shape parameters with a clear physical meaning (e.g., cut-off frequency, bandwidth, and center frequency). In this way, the number of parameters is drastically reduced and the learned feature representations are more interpretable while maintaining considerable learning flexibility. There are some attempts to learn the non-linear compression component~\cite{wang2017trainable} or learn all these components (i.e., filtering, compression, and pooling)~\cite{leaf}.

As illustrated in Figure~\ref{fig1}\,(b), these resulting neural front-ends are learned jointly along with the subsequent back-end network model, such that the representations are optimized for the downstream task at hand. However, prior neural front-ends perform feature extraction in a static fashion at the inference stage, i.e., the filter parameters are fixed once trained. That is, they can not dynamically adjust the filters conditioned on the input (e.g., varying acoustic environments) during inference. However, the evidence from psychoacoustic research has demonstrated that the mammalian cochlea performs an active mechanism exploiting metabolic energy via the outer hair cells (OHCs), which dynamically tunes its response conditioned on the input~\cite{lyon1990automatic,elliott2012cochlea}. This adaptive mechanism enables the mammalian cochlea extraordinary auditory sensitivity and selectivity, as well as a wide dynamic range of hearing~\cite{hudspeth2014integrating}. \textcolor{black}{Inspired by the mammalian auditory model, the zero-crossing with peak-amplitudes (ZCPA) model~\cite{zcpa} and power-normalized cepstral coefficients (PNCC)~\cite{PNCC} have been proposed for robust ASR.}

% \textcolor{black}{These resulting front-ends are optimized jointly with the back-end classifier to learn the representations that are customized to a specific task at hand (Figure~\ref{fig1}(a)). However, once the training is completed, the filter weights or coefficients of all these learnable front-ends are fixed at the inference stage. In other words, they do not dynamically adjust in response to varying acoustic conditions. However, psychoacoustic research has demonstrated that the outer hair cells (OHCs) of mammalian cochlea exploit metabolic energy to perform an active mechanism that dynamically adjusts its response with respect to input signal levels~\cite{lyon1990automatic,elliott2012cochlea}. It is known that such adaptive mechanisms have led to remarkable auditory sensitivity and selectivity, and a wide dynamic range of hearing~\cite{hudspeth2014integrating}.}

\textcolor{black}{In a preliminary study~\cite{wickramasinghe2023dnn}, we \textcolor{black}{introduced} a novel adaptive front-end termed Ada-FE to emulate the adaptive mechanism in the mammalian cochlea for audio spoofing detection. \textcolor{black}{Ada-FE exploits a neural feedback controller to perform adaptive inference by dynamically tuning the filter weights in a frame-wise fashion.}} The diagram of the proposed framework is shown in Figure~\ref{fig1}\,(c). In particular, our Ada-FE learns the audio representation with a set of parameterized Gabor filters that perform band-pass filtering. Ada-FE consists of one fixed Gabor filter layer and one adaptive Gabor filter layer. The adaptive Gabor filter layer exploits two parallel modules (i.e., level-dependent adaptation and adaptive neural feedback controller) to dynamically tune the filters frame by frame. The characteristic of the Gabor filter is controlled by the bandwidth and center frequency. Ada-FE tunes the shape or selectivity of filters via the quality factor (Q-factor), which is defined as the ratio of the center frequency and bandwidth.

\textcolor{black}{\textcolor{black}{Should audio front-ends be adaptive? In this paper, we conduct a systematic study on the Ada-FE across two widely used back-end classifiers and a diverse range of speech and audio tasks, involving sound event classification, non-semantic speech~\cite{shor2020towards} (keyword spotting, emotion recognition, speaker identification), and music tasks (genre classification). 
% A single-task setting is adopted for evaluation.
} Furthermore, we explore simplifying the adaptive control mechanism of the Ada-FE, and the adaptive front-end with only one adaptive filter layer to study the role of fixed Gabor filters. For the neural feedback controller, we further probe the effects of different input choices. The workflows of the Ada-FE and the simplified Ada-FE (Ada-FE-S) models are detailed in Section~\ref{sec3}.}

% In this paper, we comprehensively study the Ada-FE across two widely used back-end backbone classifiers and a diverse range of audio and speech tasks, involving acoustic scene classification, keyword spotting, emotion recognition, speaker identification, and music genre classification tasks. Furthermore, we explore simplifying the adaptive control mechanism of the Ada-FE, and the adaptive front-end with only one adaptive filter layer to study the role of fixed Gabor filters. For the neural feedback controller, we further probe the effects of different input choices. The workflows of the Ada-FE and the simplified Ada-FE (Ada-FE-S) models are detailed in Section~\ref{sec3}.

% \textcolor{blue}{In this paper, we explore further simplifying Ada-FE and more design choices with the simplified Ada-FE (Ada-FE-S). Our experiments are carried out on two widely used back-end backbone classifiers and across a diverse range of audio-related tasks, involving speech, acoustic scenes, and music tasks, to evaluate the Ada-FE and Ada-FE-S. The experimental results show that the Ada-FE and Ada-FE-S converge faster and exhibit better performance than advanced baseline systems.} 
% More notably, our adaptive front-ends demonstrate extremely excellent stability on the test set (unseen data). The workflow of the Ada-FE and Ada-FE-S models are described in Section \ref{sec3}.

\textcolor{black}{In summary, the main contribution of this work is two-fold as follows:}
\begin{itemize}
    \item \textcolor{black}{We explore the adaptive front-end (Ada-FE)~\cite{wickramasinghe2023dnn}, where a neural feedback controller is designed to dynamically tune the filter weights frame by frame at run time to emulate the active mechanism of the mammalian cochlea. Specifically, we compare Ada-FE to a range of learnable front-ends across two back-end classifiers and diverse audio and speech tasks.}
    % to emulate the active mechanism of the mammalian cochlea, which exploits a neural feedback controller to dynamically tune the filter weights in a frame-wise manner at run time. 
    \item \textcolor{black}{We further simplify the adaptive control mechanism to enable the adaptive filter layer adjusted completely by the neural feedback controller and study Ada-FE with only one adaptive filter layer. Built upon this, we probe different design choices on the input to the neural controller.}
    % We further attempt to simplify the Ada-FE by removing the control of hand-crafted level-dependent adaptation. In Ada-FE-S, the adaptive Q value is only adjusted by the neural feedback controller. Built upon Ada-FE-S, we explore more design choices on the input to the feedback controller. 
   % \item We design two modules, i.e., level-dependent adaptation and adaptive feedback controller, to dynamically learn Gabor filter layers. Comprehensive ablation studies confirm the efficacy of our design choices.
%\textcolor{black}{The comprehensive experiment results confirm the superiority of Ada-FE over the state-of-the-art front-ends in recognition accuracy and stability. Compared to Ada-FE, Ada-FE-S exhibits comparable or better performance.}
\end{itemize}

 % We comprehensively evaluate Ada-FE and Ada-FE-S across eight widely used audio/speech benchmarks and two widely used back-end backbone networks. The evaluation results show that the Ada-FE-S achieves comparable or better performance than Ada-FE. Compared to existing state-of-the-art audio front-ends, Ada-FE and Ada-FE-S demonstrate better accuracy.
 
\textcolor{black}{The remainder of this paper is structured as follows. In Section~\ref{sec2}, we first describe the related works. In Section~\ref{sec3}, we detail the proposed adaptive front-end with neural feedback control. In Section~\ref{sec4}, we present the experimental setup. Section~\ref{sec5} provides experimental results and discussions. Finally, in Section~\ref{sec6}, we conclude this paper.}

\begin{figure*}[!ht]
 \centering
  \includegraphics[width=0.994\linewidth]{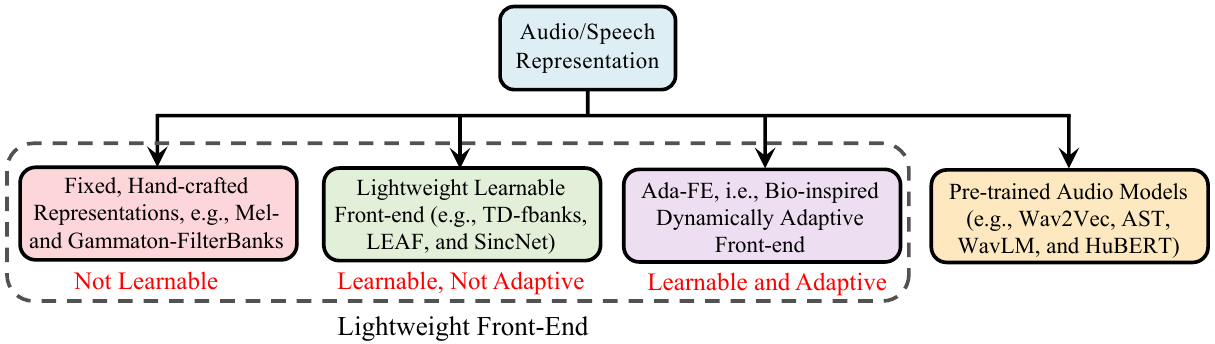}
    \caption{Illustration of audio/speech representation methods, which mainly include Fixed, hand-crafted features (e.g., Mel- and Gammaton-filterbanks), lightweight learnable front-ends (e.g., TD-fbanks, LEAF, and SincNet), our dynamically adaptive front-end (Ada-FE), and pre-trained audio models (e.g., Wav2Vec, AST, WavLM, and HuBERT).
    }
  \label{fig:sr_overview}
  % \vspace{-1.0em}
\end{figure*}

\section{Related Work}\label{sec2}

\textcolor{black}{In Figure~\ref{fig:sr_overview}, we briefly outline audio/speech representation methods, including hand-crafted fixed filterbanks, lightweight learnable front-ends (not adaptive), our dynamically adaptive front-end (Ada-FE), and audio pre-training methods.}

\subsection{\textcolor{black}{Lightweight Learnable Front-ends}} 
There have been many studies on learnable front-ends as drop-in replacements of fixed Mel-filterbanks. Some learn front-ends from low-level spectral features such as STFT. For instance, Sainath~\textit{et al.}~\cite{sainath2013learning} take the power spectral feature as the input and explore using a filterbank layer to learn the band-pass filters of Mel-filterbanks. The filterbank layer is initialized with the weights of the Mel-filters and optimized jointly with the rest of the network. Seki~\textit{et al.}~\cite{gauss} exploit the parameterized Gaussian filters to learn the band-pass filterbanks from the power spectrogram. Won~\textit{et al.}~\cite{won2020data} propose parameterized harmonic band-pass filterbanks for learnable front-end that preserves spectro-temporal locality with harmonic structures, where harmonic filters are learned from the STFT spectrogram.

Learning representations directly from the raw waveforms in an end-to-end manner has attracted great research interest, which avoids the design of hyper-parameters (such as frame length, hop length, and the typology of the window function) used to extract the appropriate low-level spectrum. In an early attempt \cite{jaitly2011learning}, Jailty and Hinton proposed pre-training a generative restricted Boltzmann machine (RBM) to learn features from raw speech signals and show that the learned features yield promising performance in phoneme recognition on the TIMIT benchmark. Later, Palaz \textit{et al.}~\cite{palaz2013estimating} proposed using a convolutional neural network (CNN) to learn representation from raw waveforms for phoneme recognition and suggest that the first layer tends to learn a set of band-pass filters. Similarly, Sainath~\textit{et al.}~\cite{sainath15_interspeech} and Hoshen \textit{et al.}~\cite{hoshen2015speech} employed CNNs for feature learning from single- and multi-channel raw waveforms respectively, where the CNN filters are initialized as Gammaton-filters and trained jointly with the deep classifier. In~\cite{dinkel2017end}, a convolutional long short-term memory deep neural network (CLDNN) is trained on raw waveforms for spoofing speech detection. Along this line of research, Zeghidour \textit{at al.}~\cite{zeghidour2018learning} proposed learnable time-domain filterbanks (TD-fbanks) for phone recognition as an approximation to Mel-filterbanks.

% and fine-tuned with the remaining CNN without constraints.

Parameterized neural filters have been explored to facilitate CNNs to discover meaningful filters. Ravanelli and Bengio~\cite{sincnet} propose the SincNet layer for speaker recognition, comprised of a set of parameterized Sinc filters that approximate band-pass filters, where the low and high cut-off frequencies are the only parameters learned from data. The recent LEAF~\cite{leaf} employs two parameterized Gabor filter layers to build a learnable front-end, with learnable filtering, pooling, and compression, demonstrating impressive results in a broad range of audio and speech tasks. The very recent works~\cite{anderson2023learnable,meng23c_interspeech} demonstrate that LEAF lacks learning as there is no substantial movement between initial filters and learned filters.

The Gabor filter, in contrast to the Sinc filter, shows characteristics similar to those of auditory filters and is optimally localized in time and frequency. This motivates us to explore parameterized Gabor filters for our adaptive front-end, where our Ada-FE learns the Q-factor to tune the filters instead of the center frequency and bandwidth used in LEAF. Additionally, unlike all prior learnable front-ends that perform feature extraction with fixed filter weights at inference time, to our knowledge, Ada-FE is the only front-end that can dynamically tune the filters frame by frame to perform adaptive inference with respect to the input.

% \textcolor{blue}{More recently, studies~\cite{anderson2023learnable,meng23c_interspeech} show that LEAF does not learn as it shows no substantial difference between learned and initialized filters, except the compression layer.} 

\subsection{\textcolor{black}{Audio Pre-training Methods}} 
\textcolor{black}{\textcolor{black}{As a parallel line of research, audio pre-training has recently fostered prominent success in audio representation learning, which involves supervised pre-training and self-supervised pre-training~\cite{audiomae,chen2023beats,byol}}. These methods typically pre-train a large neural network model on a great amount of external data to extract high-level feature representations from spectral features (e.g., AST~\cite{gong21b_interspeech}, SSAST~\cite{gong2022ssast}, and BYOL-A~\cite{byol}) or raw audio waveforms (e.g., HuBERT~\cite{hubert}, WavLM~\cite{wavlm}, and Wav2vec2~\cite{wav2vec2}).} In contrast, lightweight learnable audio front-ends only perform low-level spectral feature learning (spectral decomposition) with a small number of trainable parameters. A recent study~\cite{YadavZ22} suggests that \textcolor{black}{lightweight learnable front-ends (i.e., SincNet and LEAF) could benefit self-supervised audio pre-training. \textcolor{black}{These pre-trained models are learnable but not adaptive.}}

\section{Ada-FE: Adaptive Front-end}\label{sec3}

% \subsection{Overall Architecture}
% Prior audio front-ends perform feature extraction with fixed filer weights at inference time. Our adaptive front-end (Ada-FE) involves a neural feedback controller, in which a feed-forward neural network dynamically tunes the filter weights in a frame-by-frame fashion. 

% \textcolor{black}{In Figure~\ref{fig2}(a), we illustrate the overall architecture of our Ada-FE. Specifically, Ada-FE includes two Gabor filterbanks, where the first Gabor filterbank (purple box in Figure~\ref{fig2}(a)) employs fixed filter coefficients to maintain band-pass filtering, and the second Gabor filterbank (yellow box in Figure~\ref{fig2}(a)) consists of a set of adaptive filters. The shape of each adaptive filter is adjusted by its Q-factor defined by the ratio of the center frequency and bandwidth. The Q-factor is the only shape parameter of the neural Gabor filters learned from the data, which is produced frame by frame.}

\begin{figure*}[!ht]
% \vspace{-1.3em}
\centering
\captionsetup[sub]{font=large}
\begin{subfigure}[t]{0.95\linewidth}
% \centerline{\includegraphics[width=0.99\columnwidth]{./AdaFEv7.pdf}}
\centerline{\includegraphics[width=0.99\columnwidth]{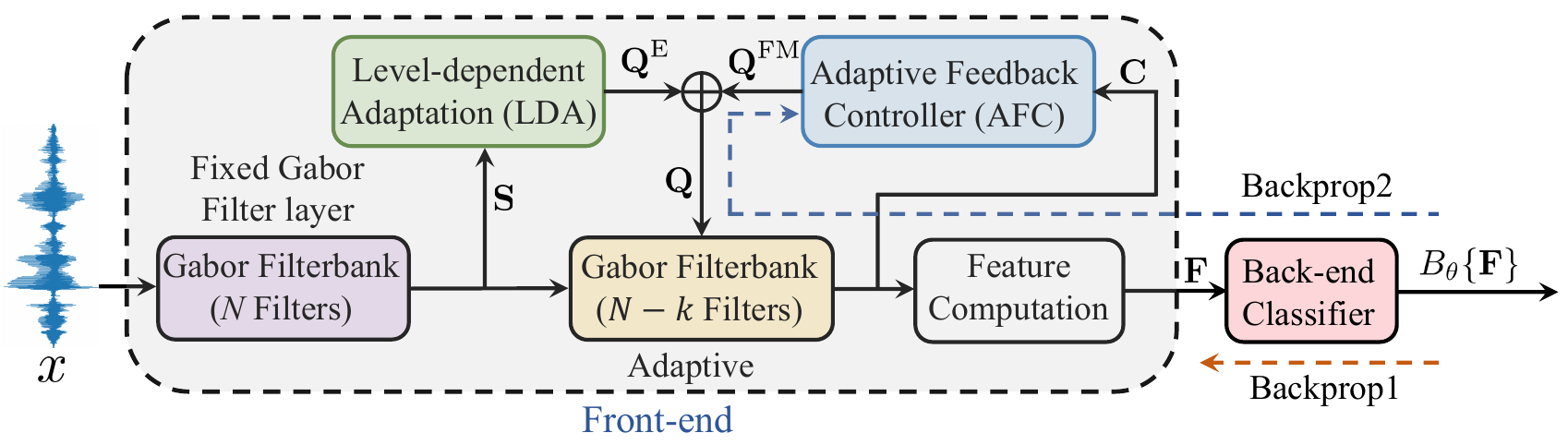}}
% \vspace{-0.5em}
\caption{}
% \caption{The adaptive front-end (Ada-FE) is optimized jointly with the back-end classifier.}
\label{fig2:1}
\end{subfigure}
\begin{subfigure}[t]{0.95\linewidth}
% \centerline{\includegraphics[width=0.99\columnwidth]{./AdaFES_V5.pdf}}
\centerline{\includegraphics[width=0.99\columnwidth]{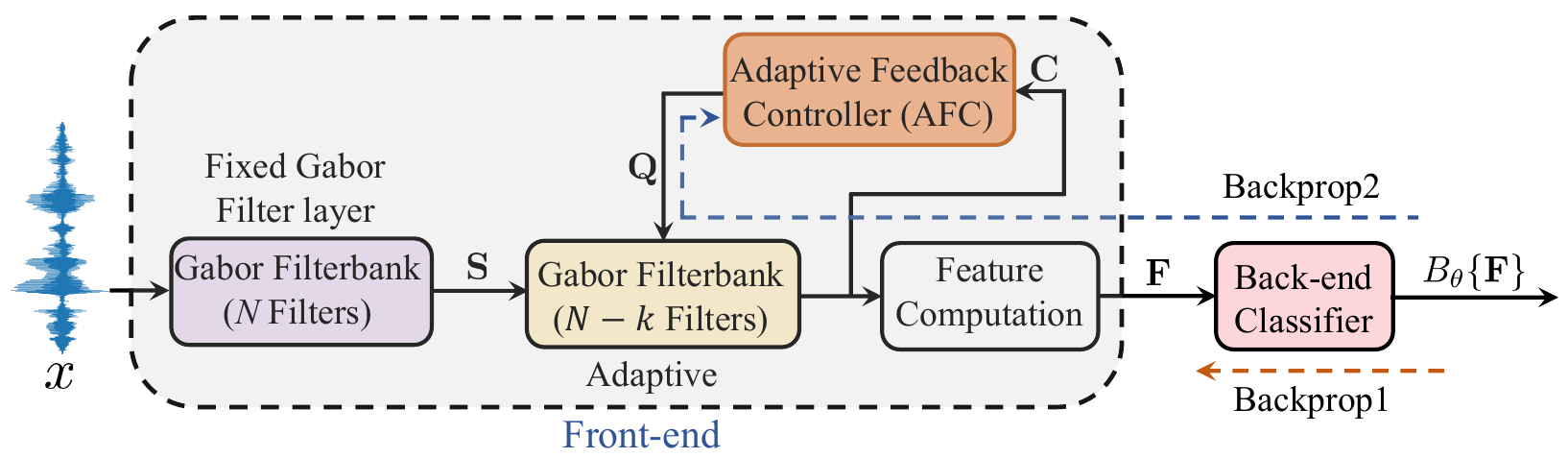}}
% \vspace{-0.5em}
\caption{}
% \caption{The simplified adaptive front-end (Ada-FE-S) is optimized jointly with the back-end classifier.}
\label{fig2:2}
\end{subfigure}
\caption{\textcolor{black}{Illustrations of (a) the overall diagram of the adaptive front-end (Ada-FE)~\cite{wickramasinghe2023dnn} and \textcolor{black}{(b) the simplified adaptive front-end (Ada-FE-S), where the hand-crafted level-dependent adaptation function module is removed and the adaptive Q value is completely controlled by the neural adaptive feedback controller (orange box).}}}
\vspace{-0.1em}
\label{fig2}
\end{figure*}

% \textcolor{black}{In Figure~\ref{fig2}\,(a), we illustrate the overall architecture of our Ada-FE. In particular, Ada-FE contains one fixed Gabor filter layer (purple box in Figure~\ref{fig2}\,(a)) and one adaptive Gabor filter layer (yellow box in Figure~\ref{fig2}\,(a)). The fixed Gabor filter layer maintains band-pass filtering with fixed weights. The adaptive filter layer dynamically tunes the filters by two parallel modules (via the Q-factor), i.e., level-dependent adaptation (LDA, green box) and an adaptive feedback controller (AFC, blue box). The Q-factor, which is defined by the ratio of the center frequency and bandwidth, is the only shape parameter of the adaptive Gabor filters learned from the data. For each adaptive filter, its Q value is attained by the summation of $Q^{\text{E}}$ and $Q^{\text{FM}}$ calculated from the LDA and the AFC modules respectively: $Q\{x\}\!=\!Q^{\text{E}}\! +\! Q^{\text{FM}}$. The LDA module employs a hand-crafted piecewise function to calculate the Q value according to the energy of each subband, which is described in detail in Section~\ref{sec3.2}. The AFC exploits a two-layer fully-connected neural network to dynamically tune the Q factors in a frame-wise manner, where the Q factors for the current time frame are calculated from the last time frame, behaving like a recurrent neural network.} 

In Figure~\ref{fig2}\,(a), we illustrate the overall architecture of our Ada-FE~\cite{wickramasinghe2023dnn}, which contains one fixed Gabor filter layer (purple box in Figure~\ref{fig2}\,(a)) and one adaptive Gabor filter layer (yellow box in Figure~\ref{fig2}\,(a)). \textcolor{black}{There is no other processing between these two Gabor filterbanks.} The fixed Gabor filter layer maintains band-pass filtering with fixed weights. The adaptive filter layer dynamically tunes the filters by two parallel modules (via the Q-factor), i.e., level-dependent adaptation (LDA, green box) and an adaptive feedback controller (AFC, blue box). The Q-factor, which is defined by the ratio of the center frequency and bandwidth, is the only shape parameter of the adaptive Gabor filters learned from the data. For each adaptive filter, its Q value is attained by the summation of \textcolor{black}{$\textbf{Q}^{\text{E}}$ and $\textbf{Q}^{\text{FM}}$ calculated from the LDA and the AFC modules respectively: $\textbf{Q}\!=\!\textbf{Q}^{\text{E}}\! +\! \textbf{Q}^{\text{FM}}$. The LDA module employs a hand-crafted piecewise function to calculate the Q value according to the energy of each subband, which is described in detail in Section~\ref{sec3.2}. The AFC exploits a two-layer fully-connected neural network to dynamically tune the Q-factors in a frame-wise manner, where the Q-factors for the current time frame are calculated from the last time frame, behaving like a recurrent neural network.} 

\textcolor{black}{In addition to the broader question of whether front-ends should be adaptive, in this paper we also explore if the role of the feedforward level-dependent adaptation can be subsumed into the adaptive neural feedback controller, i.e., AFC. }
%in this paper, we are particularly interested in learning the adaptive Q value with only the adaptive neural control module. i.e., AFC. 
To this end, we attempt to further simplify Ada-FE by removing the hand-crafted LDA module. Figure~\ref{fig2}\,(b) shows the architecture of the simplified Ada-FE (Ada-FE-S). In this way, the adaptive Q value is completely adjusted by the neural AFC. Exploiting the parameterized Gabor filters, where only the Q-factor is the tunable parameter to adjust the filter response, can place the filters on an appropriate frequency scale and still allow considerable learning flexibility. Moreover, this solution also mitigates the overfitting and stability issues encountered by training unconstrained filters~\cite{sincnet,leaf}.

\textcolor{black}{The neural AFC is jointly trained with the parameters of the back-end classifier using gradient-based optimization methods. However, this joint training is not straightforward since the direction of information flow through the neural AFC runs counter to the rest of the neural network. In particular, the input to the controller is from a later stage of the network while the output of the controller feeds to an earlier stage of the network (i.e., information flows from the output of the neural network toward the input through the feedback controller). Consequently, the paths of gradient flow for backpropagation are also split into two and are depicted as Backprop1 and Backprop2 in Figure~\ref{fig2}. This joint optimization solution allows the AFC to be updated without the need for labeled Q-factor values, instead updating with the objective of reducing the classification error.}

\textcolor{black}{During training, the adaptive controller is expected to dynamically tune the filters to better extract spectral features for downstream tasks. Since the features that need to be enhanced and the factors that affect this vary across time. i.e., certain bands need to be enhanced at certain times, the Q-factor value must be dynamically adaptive. Each component of the Ada-FE architecture is detailed in the following Sections.}

% \begin{figure}[!htbp]
% % \vspace{-1.3em}
% \centering
% \begin{subfigure}[t]{0.85\columnwidth}
% \centerline{\includegraphics[width=0.97\columnwidth]{Gabortime.pdf}}
% \caption{Time impulse responses of Gabor filters}
% \label{fig3:1}
% \end{subfigure}
% \begin{subfigure}[t]{0.88\columnwidth}
% \centerline{\includegraphics[width=0.99\columnwidth]{Gaborfreq.pdf}}
% \caption{Frequency responses of Gabor filters}
% \label{fig3:2}
% \end{subfigure}
% \caption{Illustrations of the time impulse and frequency responses of three Gabor filters, where $f_{c}\!=\!\left\{1000, 2500, 3500\right\}$ and $Q\!=\!\left\{1.0, 1.5, 2.5\right\}$, respectively.}
% % \vspace{-1.0em}
% \label{fig3}
% \end{figure}

\subsection{Gabor Filter Layer}

\textcolor{black}{\textcolor{black}{Let $x$ denote the input waveform which is filtered frame-wise with the first Gabor filter layer comprising} of $N$ fixed band-pass filters, $\textbf{W}^{1}\!=\!\{W^{1}_{1}, W^{1}_{2}, ...,W^{1}_{N}\}$. This produces a feature map with $N$ feature channels, denoted as $\textbf{Y}\in\mathbb{R}^{T\!\times\! N\times\! F}$, where $T$ and $F$ represent the number of time frames and frame length, respectively. \textcolor{black}{Following} this, a $k^{th}$ order spatial differentiation operation~\cite{WickramasingheA19} is applied to $\textbf{Y}$. For uniformly spaced filters, first-order spatial differentiation is implemented by taking the difference between two adjacent feature channels, and higher-order spatial differentiations can be implemented by repeating this operation. The spatial differentiation has been shown to increase the sharpness of filter transition bands and reduce their overlap, which reduces the correlation between adjacent feature coefficients~\cite{wickramasinghe2021replay}.}

\textcolor{black}{The spatial differentiation produces a feature map denoted as $\textbf{S}\!=\!\{\textbf{S}_{i}\}$, where \textcolor{black}{$\mathbf{S}_{i}\!\in\!\mathbb{R}^{(N-k)\!\times\!F}$} represents the feature map for $i^{th}$ time frame, with $i\!\in\!\{1,..., T\}$. The feature map $\mathbf{S}$ is then filtered frame-wise by the second Gabor filter layer, comprising \textcolor{black}{$N\!-\!k$} adaptive band-pass filters, \textcolor{black}{$\textbf{W}^{2}\!=\!\{W^{2}_{1}, W^{2}_{2}, ...,W^{2}_{N-k}\}$.} Here, $W^{1}_{i}\!\in\!\mathbb{R}^{P}$ and $W^{2}_{i}\!\in\!\mathbb{R}^{P}$ denote the $i^{th}$ filters in the first and the second Gabor filter layers, respectively, where $P$ is filter length. Each output frequency channel from the first layer is processed by the corresponding channel in the second layer. The center frequencies of second-layer filters are calculated to match the center frequency shift produced by spatial differentiation operation. 
}

\begin{figure}[!tbp]
% \vspace{-1.3em}
\centering
\begin{subfigure}[t]{0.85\columnwidth}
\centerline{\includegraphics[width=0.97\columnwidth]{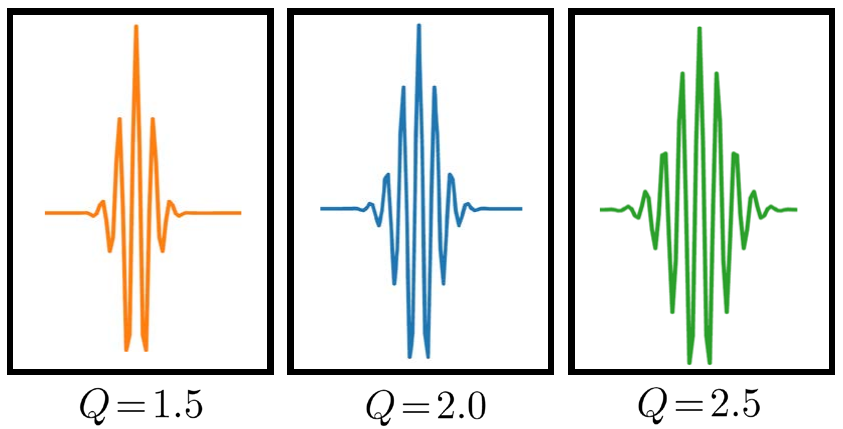}}
\caption{Time impulse responses of Gabor filters}
\label{fig3:1}
\end{subfigure}
\begin{subfigure}[t]{0.85\columnwidth}
\centerline{\includegraphics[width=0.99\columnwidth]{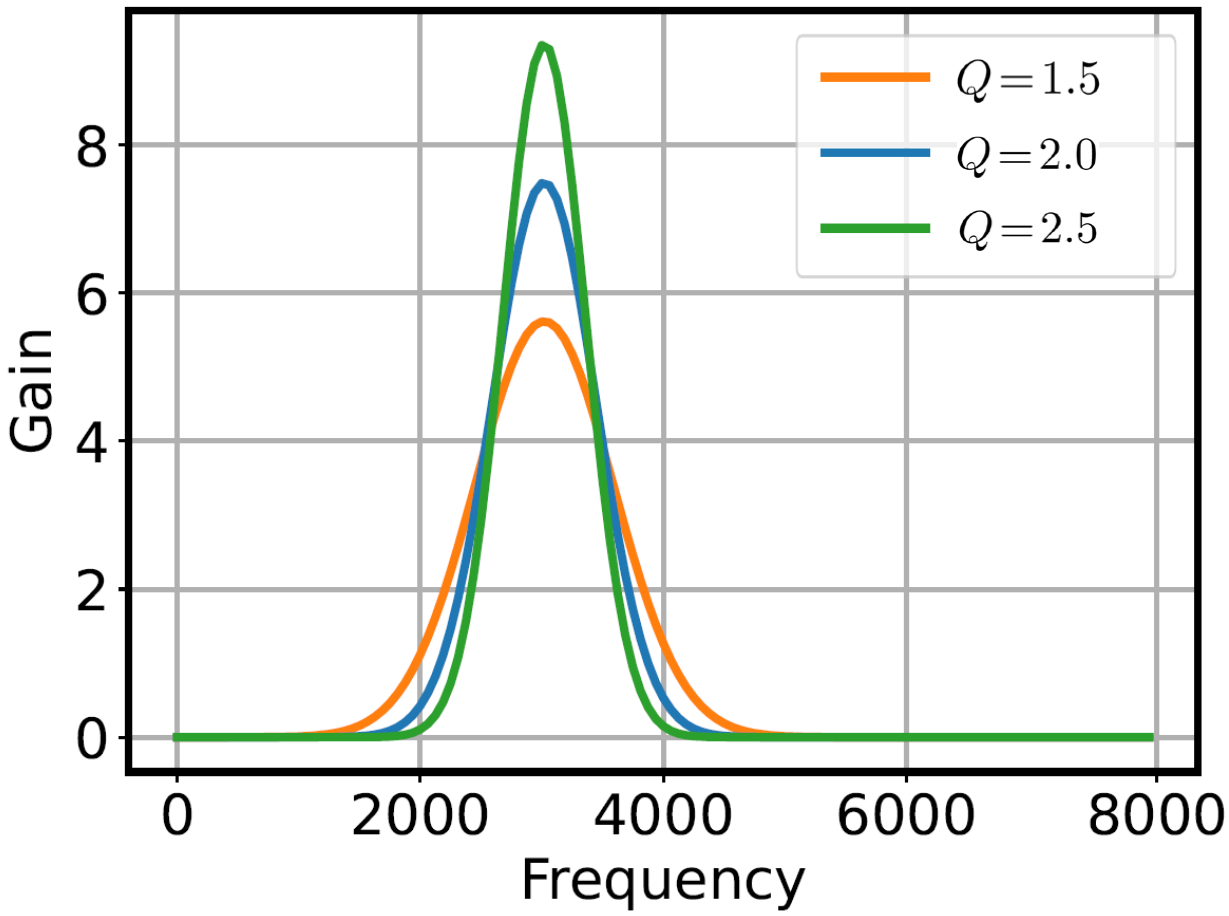}}
\caption{Frequency responses of Gabor filters}
\label{fig3:2}
\end{subfigure}
\caption{Illustrations of the time impulse and frequency responses of three Gabor filters, \textcolor{black}{where $f_{c}\!=\!3000$ and $Q\!=\!\left\{1.5, 2.0, 2.5\right\}$, respectively.}
% $f_{c}\!=\!\left\{1000, 2500, 3500\right\}$ and $Q\!=\!\left\{1.0, 1.5, 2.5\right\}$, respectively.
}
\vspace{-1.0em}
\label{fig3}
\end{figure}

\textcolor{black}{At training time, the first Gabor filter layer, $\textbf{W}^{1}$, is fixed, while the second Gabor filter layer, $\textbf{W}^{2}$, is adaptive. Gabor filter provides a straightforward relationship between its Q-factor and filter response or shape. The time impulse response of the chosen Gabor filter is defined as \cite{gaborfilter}:
% \vspace{-0.5em}
\begin{equation}
W[z]=e^{-(b z)^2} \cos \left(\Omega_c z\right)
\label{eq1}
\end{equation}
where $\Omega_{c}\!=\!2\pi\frac{f_{c}}{f_{s}}$, $b\!=\!{\sqrt{2\pi}}\frac{BW}{2f_{s}}$, and $z\!\in\!\{1,...,\textcolor{black}{P}\}$. $f_{c}$ and $f_{s}$ denote the center frequency and the sampling frequency, respectively, and $BW$ denotes the 3-dB bandwidth, given as $BW\!=\!{\frac{f_{c}}{Q}}$.} \textcolor{black}{Let $\Omega\!=\!\Omega_{c}$, the relationship between the Q-factor and its frequency response at the center frequency is given by 
\begin{equation}
  \left|W\left(\Omega_c\right)\right|=\frac{\sqrt{2} \pi \cdot Q}{\Omega_c}(1+e^{-8 \pi Q^2}).  
\end{equation}
When $Q\!\geq\!0.5$, the factor $e^{-8 \pi Q^2}$ is very small and we obtain $\left|W\left(\Omega_c\right)\right| \approx \frac{\sqrt{2} \pi \cdot Q}{\Omega_c}$. \textcolor{black}{In Figure~\ref{fig3}\,(a) and (b), we illustrate the time impulse responses and frequency responses of three Gabor filters, respectively, \textcolor{black}{where $f_{c}\!=\!3000$ and $Q\!=\!\{1.5,2.0,2.5\}$. It visualizes how the selectivity (gain) and sensitivity (bandwidth) of filters are adjusted via the Q-factor.}}}
% with $f_{c}\!=\!\{1000,2500,3500\}$ and $Q\!=\!\{1.0,1.5,2.5\}$. It visualizes how the selectivity (gain) and sensitivity (bandwidth) of filters are adjusted via the Q-factor.

\textcolor{black}{As shown in Figure~\ref{fig2}\,(a), the implementation of the adaptive Q-factor in Ada-FE involves level-dependent adaptation (LDA) and neural adaptive feedback controller (AFC). The adaptive controller dynamically tunes the Q factors frame by frame. At frame $t-1$, with the input as \textcolor{blue}{$\mathbf{S}_{t-1}\!\in\!\mathbb{R}^{(N-k)\times F}$}, LDA yields the Q value $\mathbf{Q}_{t-1}^{\text{E}}\!\in\!\mathbb{R}^{(N-k)\times1}$. The AFC takes the input as \textcolor{blue}{the output of the adaptive filter layer $\mathbf{C}_{t-1}\!\in\!\mathbb{R}^{(N-k)\!\times\!F}$} and produces the Q value $\mathbf{Q}_{t-1}^{\text{FM}}\!\in\!\mathbb{R}^{(N-k)\times 1}$. The adaptive Q value for $t^{th}$ frame is updated as $\mathbf{Q}_{t}\!=\!\mathbf{Q}_{t-1}^{\text{E}}+\mathbf{Q}_{t-1}^{\text{FM}}$ for \textcolor{black}{$\!N-\!k$} filters.}

\subsection{Adaptive Controller} \label{sec3.2}

% \textcolor{black}{In Ada-FE (Figure~\ref{fig2}\,(a)), the implementation of the adaptive Q-factor involves level-dependent adaptation (LDA) and neural adaptive feedback controller (AFC).}

\begin{figure}[!ht]
 \centering
  \includegraphics[width=0.94\linewidth]{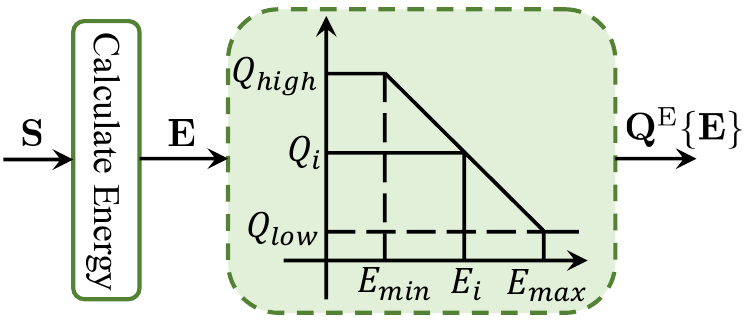}
    \caption{Illustration of the LDA module.}
  \label{fig4}
  \vspace{-0.5em}
\end{figure}

\textbf{Level-Dependent Adaptation}. In our previous study~\cite{WickramasingheA19}, we propose a feed-forward adaptive front-end consisting of two filterbanks for spoofed speech detection, where an LDA module is employed to provide variable selective gain via Q value according to the input levels. As shown in Figure \ref{fig4}, the energy $\mathbf{E}$ corresponding to each subband or frequency channel was calculated over the spatially differentiated filter output $\mathbf{S}$. The Q value $\textbf{Q}^{\text{E}}\{\mathbf{E}\}$ is low for a high input level $\mathbf{E}$, resulting in a filter with decreased gain while increasing the bandwidth, which enables the filter to be less selective and compresses the input signal. \textcolor{black}{In contrast, $\mathbf{Q}^{\text{E}}$ is high for a low input level, which results in a filter with increased gain and reduced bandwidth. It enables the filter to be more selective.}

\begin{figure}[!t]
 \centering
    \includegraphics[width=0.99\linewidth]{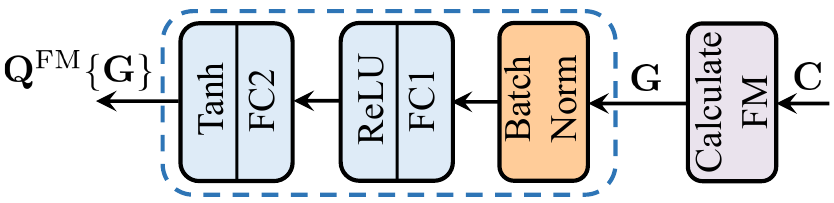}
  % \vspace{-0.5em}
    \caption{Illustration of the AFC module.}
  \label{fig5}
  % \vspace{-0.9em}
\end{figure}

\textbf{Neural Adaptive Feedback Controller}. \textcolor{black}{As illustrated in Figure~\ref{fig5} (blue dotted box), the neural AFC takes the frame-averaged frequency modulation (FM) component $\mathbf{G}$ of one given subband as the input~\cite{wickramasinghe2021replay}. The input is first processed by a batch normalization (BN) operation and then flows into a two-layer fully-connected (FC) network \textcolor{black}{of size \textcolor{black}{$N-k$}}, with ReLU and Tanh activation respectively. The resulting Q value $\mathbf{Q}^{\text{FM}}$ is added with the Q value $\mathbf{Q}^{\text{E}}$ that depends on input energy (relating to amplitude modulation (AM)) to obtain the final Q value to update the Gabor filters. This is motivated by evidence in psychoacoustic research, which shows that the FM component is complementary to the AM component within the auditory system and captures spectral information for accurate signal-noise separation for speech perception~\cite{nie2004encoding}. Additionally, phase information carried in the FM component reflects reverberation effects and spectral envelope changes due to channel characteristics. The FM component $\mathbf{G}$ is computed from the output of the second Gabor filter layer. For FM calculation, we employ the spectral centroid deviation method~\cite{fm2018}. In this paper, we explore different design choices for the input to the AFC module, which will be described in Section~\ref{sec:5.2}}

\textcolor{black}{Since an increase of Q value increases the ringing in the filter, some constraints should be considered to ensure the ringing is not too high as it may degrade the performance. To limit the maximum Q values learned by FC1 in a differentiable manner, we employ a scaled and shifted version of an FC layer (FC2) with a diagonal weight matrix and biases followed by the Tanh activation~\cite{wickramasinghe2023dnn}.} \textcolor{black}{Frame-wise processing of the whole utterance is followed by the feature computation module and a back-end classifier.}

% \textcolor{black}{Since an increase of Q value increases the ringing in the filter, some constraints should be considered to ensure the ringing is not too high as it may degrade the performance. To limit the maximum Q values learned by FC1 in a differentiable manner, we employ a scaled and shifted version of an FC layer (FC2) with a diagonal weight matrix and biases followed by the Tanh activation~\cite{wickramasinghe2023dnn}.} \textcolor{black}{The adaptive controller dynamically tunes the Q factors frame by frame. Frame-wise processing of the whole utterance is followed by the feature computation module and a back-end classifier.}

\subsection{Feature Computation}

The adaptive Gabor filter layer is followed by a feature calculation operation, which involves subband energy and spectral envelop centroid magnitude. \textcolor{black}{Information about subband energy can be discerned from the mean magnitude of the Fourier transform of the subband signal.}
%Subband energy can be viewed as the mean magnitude of the Fourier transform of a subband signal. 
The spectral envelope centroid magnitude (CM), which is defined as the weighted average magnitude of the spectral envelope of a given subband signal, is introduced for spoofing attack detection~\cite{8683693}. In this study, we divide the full-band subband spectral envelope (from 0 to 8 kHz) into five octaves, and the CM feature is extracted from each octave. CM from $j^{th}$ octave of the subband signal of a given frame is
% In this study, we divide the full-band subband spectral envelope (from 0 to 8 kHz) into five octaves, and the CM feature is extracted from each octave as the equation (8) in~\cite{8683693}. CM from $j^{th}$ octave of the subband signal of a given frame is calculated as: 
% \vspace{-0.6em}
\begin{equation}
\text{CM}_j=\frac{\sum_{f=f_l^{(j)}}^{f_u^{(j)}} f\cdot|E[f]|}{\sum_{f=f_l^{(j)}}^{f_u^{(j)}} f}
% \vspace{-0.7em}
\end{equation}
\textcolor{black}{where $|E[f]|$ denotes the spectral envelope, $f$ denotes the frequency, and $f_{l}^{(j)}$ and $f_{u}^{(j)}$ denote the lower and upper-frequency bounds of the $j^{th}$ octave  respectively.}

\input{dataset.tex}

\section{Experimental Setup}\label{sec4}

\subsection{Benchmarks and Datasets}
% \textcolor{black}{\textcolor{black}{\textcolor{blue}{To perform a comprehensive investigation, audio front-ends should be evaluated across a diverse and wide range of audio signals~\cite{gong21b_interspeech,gong2022ssast,byol}. To this end, we evaluate the front-ends on eight popular audio and speech datasets, across acoustic scene classification, non-semantic speech task (speech emotion classification, keyword spotting, speaker identification), and music tasks (genre classification).}} For all the tasks, we report classification accuracy as the evaluation metric. Table~\ref{dataset} shows the eight datasets for evaluation and the descriptions are given below.}
\textcolor{black}{\textcolor{black}{\textcolor{black}{To perform a comprehensive investigation, audio front-ends should be evaluated across a diverse and broad range of audio signals. To this end, we evaluate the audio front-ends on eight audio and speech classification benchmarks that are widely used in previous studies~\cite{chen2023beats,audiomae,gong21b_interspeech,gong2022ssast,byol}, including sound event classification, non-semantic speech~\cite{byol,shor2020towards}, and music tasks (genre classification). The speech tasks involve keyword spotting, emotion classification, and speaker identification, which belong to three different aspects of speech~\cite{superb}, i.e., content, paralinguistics, and speaker, respectively.}} For all the tasks, we report classification accuracy as the evaluation metric. Table~\ref{dataset} shows the eight datasets for evaluation and the descriptions are given below.}
\begin{itemize}
\item \textcolor{black}{\textbf{Environmental Sound Classification} (ESC-50) dataset \cite{esc50} is employed for the sound event classification task. \textcolor{black}{ESC-50 comprises $2\,000$ environmental audio recordings grouped into 50 classes, where each recording is 5 seconds long with a sampling rate of $44.1$ kHz} and labeled with only one class. We perform the standard five-fold cross-validation to evaluate performance and report the accuracy~\cite{gong21b_interspeech}.}

\item \textcolor{black}{\textbf{GTZAN} We use the GTZAN dataset~\cite{gtzan} for the music genre classification task. \textcolor{black}{GTZAN comprises $1\,000$ single-label music clips of 10 genre classes, and each audio clip is 30 seconds long with a sampling rate of $22.05$ kHz.} We follow the fault-filtered split setting provided in~\cite{tmm15,byol} to evaluate models and report the accuracy.}

\item \textcolor{black}{\textbf{Free Music Archive} (FMA) dataset~\cite{fma} is employed for music information analysis. In this paper, we also use the FMA dataset for the music genre classification task. \textcolor{black}{Here, we use the small subset of FMA (FMA-S), which consists of $8\,000$ 30-second-long audio clips (sampled at $44.1$ kHz) of 8 genres ($1\,000$ clips per genre).} We follow the standard split provided in the dataset to evaluate models and report the accuracy.}

\item \textcolor{black}{\textbf{CREMA-D} \textcolor{black}{We use the CREMA-D dataset~\cite{cremad} for the speech emotion task, comprising $7\,442$ audio clips (sampled at $16$ kHz) from 43 female and 48 male speakers.} For CREMA-D, we follow the split setting employed in~\cite{byol}, which uses the 7:1:2 split to assign speakers into training, validation, and testing subsets (without duplication) to evaluate models and report accuracy.
}

\item \textcolor{black}{\textbf{IEMOCAP} We also employ the commonly used IEMOCAP dataset \cite{iemocap} for the speech emotion recognition task, which comprises approximately 12 hours of speech clips from 10 speakers. For IEMOCAP, we adopt the SUPERB evaluation benchmark~\cite{superb} that follows the conventional evaluation protocol, where the unbalanced emotion classes are dropped to leave the four classes \textcolor{black}{(angry, happy, sad, and neutral)} with a similar amount of speech clips. The standard five-fold cross-validation evaluation is performed and the accuracy is reported. \textcolor{black}{All audio samples are at a sampling rate of $16$ kHz.}}

\item \textcolor{black}{\textbf{Speech Commands V2 (SPC-V2)} We use the Google SPC-V2 dataset \cite{GSCv2} for the keyword spotting task. SPC-V2 contains a total of $105\,829$ 1-second-long utterances from $2\,618$ speakers, which are recorded at a sampling rate of 16 kHz and labeled with one of the 35 common speech commands. Following the standard split setting provided in the dataset~\cite{GSCv2}, we split the dataset into training, validation, and test sets, with $84\,843$, $9\,981$, and $11\,005$ utterances, respectively. Here, we focus on the 35-class classification task and repost the accuracy, no class re-balancing operation is applied during testing and training.}

\item \textcolor{black}{\textbf{Speech Commands V1 (SPC-V1)} We also employ the Google SPC-V1 \cite{GSCv2} for the keyword spotting task. \textcolor{black}{SPC-V1 contains a total of $64\,721$ 1-second-long utterances from $1\,881$ speakers, which are recorded at a sampling rate of $16$ kHz and labeled with one of the 30 common speech commands.} We follow the official split setting provided in the dataset~\cite{GSCv2} and focus on the 30-class classification task. We report the accuracy, no class re-balancing operation is applied during testing and training.}

\item \textcolor{black}{\textbf{VoxCeleb1} We employ the VoxCeleb1 dataset~\cite{voxceleb1} for the speaker identification (speaker ID) task, which consists of a total of $153\,516$ recordings from $1\,251$ speakers. All the speech signals are sampled at 16 kHz. Following the default identification split provided by the VoxCeleb1, the dataset is split into a training set with $138\,316$ utterances, a development set with $6\,904$ utterances, and a test set with $8\,251$ utterances, respectively. All the recordings of the three sets are from the same $1\,251$ speakers.}
\end{itemize}

\subsection{Implementation Details}

\textcolor{black}{Following~\cite{chen2023beats,audiomae,gong21b_interspeech,gong2022ssast,byol}, the raw audio signals are resampled to $16$ kHz.} We segment audio and speech recordings into 11 ms long non-overlapping time frames for frame-wise \textcolor{black}{adaptation}. For the fixed Gabor filter layer, we employ $N\!=\!40$ normalized Gabor filters (in Hz frequency scale) with a length of $L\!=\!150$ samples. The filtered signal is then passed through a first-order ($k\!=\!1$) spatial differentiation operation~\cite{8683693}, which generates \textcolor{black}{$N\!-\!1=39$} output channels for each time frame. They are then fed into the adaptive Gabor filter layer comprised of \textcolor{black}{$N-1\!=\!39$} filters to tune Q values on a frame-by-frame basis.

% The fixed Gabor filterbank contains $N\!=\!40$ normalized Gabor filters (in Hz frequency scale) of length $L\!=\!150$ coefficients. The first-order ($k\!=\!1$) spatial differentiation is then applied to the filtered signal. The resulting 39 ($M\!=\!N\!-\!k$) output channels per framed signal are fed into the adaptive Gabor filter layer containing $M\!=\!39$ filters.

\textcolor{black}{In our experiments, we compare our Ada-FE and Ada-FE-S with recent TD-fbanks and LEAF, where an independent front-end is trained jointly with the back-end for each task.} \textcolor{black}{Since LEAF demonstrates state-of-the-art results over Mel-filterbanks and prior learnable front-ends in diverse audio and speech tasks, we mainly focus on the comparison to LEAF.} For the TD-fbansks baseline, here we report the best results listed in the work~\cite{leaf}. We use the official code implementation of the LEAF baseline to run all the experiments. We follow LEAF~\cite{leaf} evaluating front-ends with the same common network architecture, which comprises a front-end followed by a back-end classifier. To ensure fair comparisons, our comparative study keeps the same experiment settings.

% Our experiments employ the TD-fbanks and LEAF as the baselines. Here, we mainly focus on the comparison of the Ada-FE and Ada-FE-S with LEAF, since LEAF demonstrates state-of-the-art performance in diverse audio and speech tasks. For TD-fbansks, the best results reported in \cite{leaf} are listed here. Our experiments adopt the official code implementation for the LEAF baseline model and keep the same experiment settings for a fair comparison. Consistent with LEAF, we employ the same common network architecture, which consists of a front-end and a back-end classifier. 

\textbf{EfficientNets}~\cite{efficientnet} is a family of CNNs, that demonstrate substantial superiority in accuracy and efficiency over previous convolutional network architectures. The superiority mainly comes from two design choices: 1) EfficientNets scale up neural networks by all three dimensions (i.e., input resolution, width, and depth) with a compound coefficient, which performs better than scaling only one dimension; 2) EfficientNets use an efficient building block, i.e., mobile inverted bottleneck convolution (MBConv) with squeeze-and-excitation optimization~\cite{senet}. Following LEAF paper~\cite{leaf}, we use EfficientNet-B0, i.e., a basic version of EfficientNets, to perform the back-end classifier.

% and involves 5.3M parameters

\textbf{MobileNetV2}~\cite{mobilenetv2} is a family of highly efficient mobile convolutional networks. The core layer module of the MobileNetV2 is the inverted residual with linear bottleneck, which allows one to significantly reduce memory requirement at inference time. To conduct comprehensive comparison experiments, we also study LEAF and Ada-FE (as well as Ada-FE-S) on the MobileNetV2 backbone network. In our experiments, we employ the MobileNetV2-100 as the back-end classifier\footnote{The code implementation for MobileNetV2-100 is adopted from PyTorch Image Models (timm) library, which is available here \url{https://github.com/huggingface/pytorch-image-models}.}.

% MobileNetV2 is a family of highly efficient mobile convolutional network models. The core layer module of MobileNetV2 is the inverted residual with linear bottleneck, which allows one to significantly reduce memory requirement at inference time. In this study, we employ the MobileNetV2-100\footnote{The code implementation for MobileNetV2-100 is adopted from PyTorch Image Models (timm) library, which is available here \url{https://github.com/huggingface/pytorch-image-models}.} as the back-end classifier.

\textbf{Training details}. We employ the cross-entropy as the loss function and the~\textit{Adam} optimizer for the gradient optimization, with the parameters~\cite{Adam}, i.e., $\!\beta_{1}\!=\!0.9$, $\!\beta_{2}\!=\!0.98$, and $\epsilon\!=\!10^{-9}$. The models are trained for 150 epochs, with a batch size of 64 utterances. The initial learning rate is set to $1.0\!\times\!10^{-4}$, with a weight decay of $1.0\!\times\! 10^{-4}$. To handle the variable duration of the input audio and speech recordings, following LEAF~\cite{leaf} we employ randomly sampled 1-second clips for training. At the inference stage, we use the entire recordings for prediction, dividing them into consecutive non-overlapping 1-second clips and taking the average of the output logits over clips to attain the final prediction or classification results. Moreover, as a more common design choice in speaker identification~\cite{voxceleb1}, we also train models on randomly sampled 3-second segments and employ the same method to perform inference. It is worth noting that no data augmentation is applied to the raw audio waveforms at training time for fair comparisons. \textcolor{black}{We repeat each experiment three times and report the mean results.}

\section{Experimental Results}\label{sec5}

\subsection{Q-Value Analysis}

% is consistent with the response of cochlear to varying input levels, 

% which is consistent with those shown in Figures 4-6.

% \begin{figure}[!htbp]
% % \vspace{-1.3em}
% \centering
% \begin{subfigure}[t]{0.50\columnwidth}
% \centerline{\includegraphics[width=1.0\columnwidth]{E_SPCV2.pdf}}
% \caption{The Energy curve}
% \label{fig1:1}
% \end{subfigure}
% \begin{subfigure}[t]{0.48\columnwidth}
% \centerline{\includegraphics[width=0.98\columnwidth]{Q_SPCV2.pdf}}
% \caption{The Q value curve}
% \label{fig1:2}
% \end{subfigure}
% \caption{Illustration of how (a) energy and (b) Q value change frame by frame for the Gabor filter with the center frequency of $3.27$ KHz at inference time. Ada-FE infers one 1-second speech utterance on the keyword spotting (SPC-V2).}
% \vspace{-0.5em}
% \label{figQ1}
% \end{figure}

\begin{figure}[!h]
% \vspace{-1.3em}
\centering
\begin{subfigure}[t]{0.49\columnwidth}
\centerline{\includegraphics[width=1.0\columnwidth]{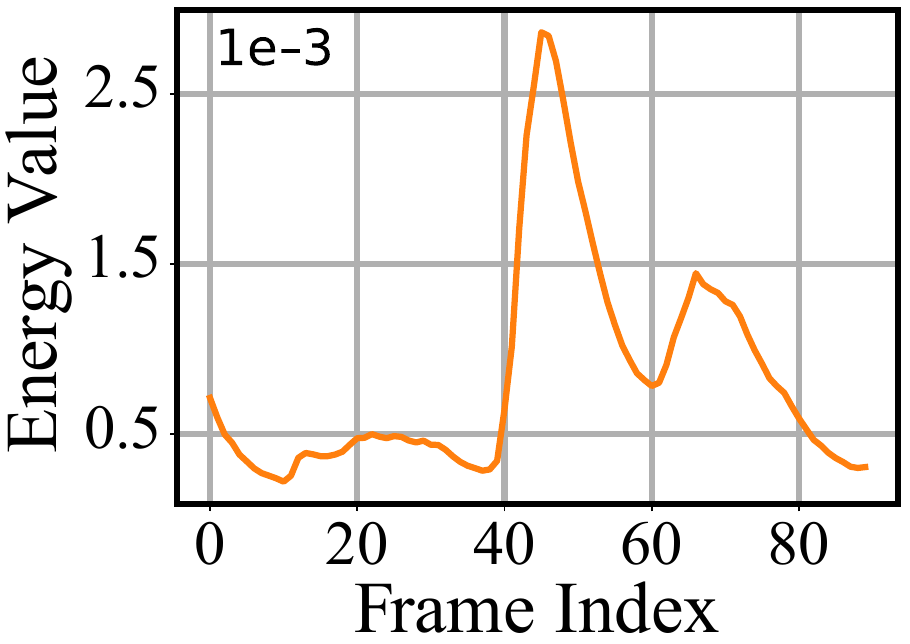}}
\caption{The Energy curve}
\label{fig1:1}
\end{subfigure}
\begin{subfigure}[t]{0.49\columnwidth}
\centerline{\includegraphics[width=0.98\columnwidth]{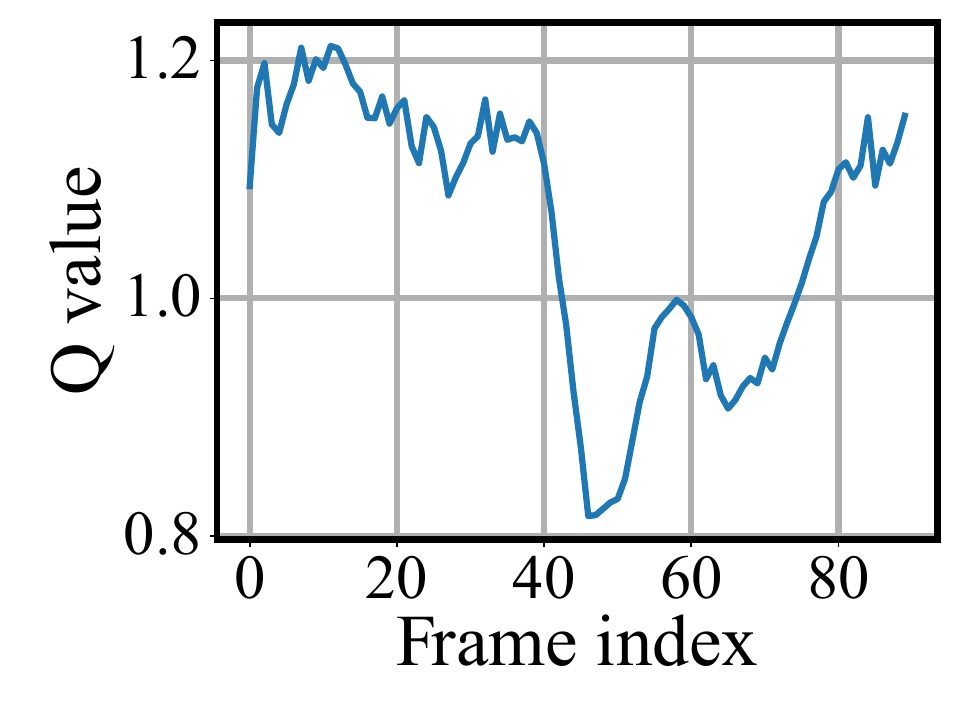}}
\caption{The Q value curve}
\label{fig1:2}
\end{subfigure}
\caption{\textcolor{black}{Illustration of how (a) energy and (b) Q value change frame by frame for the Gabor filter with a center frequency of $1.12$ kHz (low). Ada-FE infers one 1-second speech segment on the speaker ID task (VoxCeleb1).}}
\vspace{-0.5em}
\label{figQ1}
\end{figure}

\begin{figure}[!h]
% \vspace{-1.3em}
\centering
\begin{subfigure}[t]{0.505\columnwidth}
\centerline{\includegraphics[width=1.0\columnwidth]{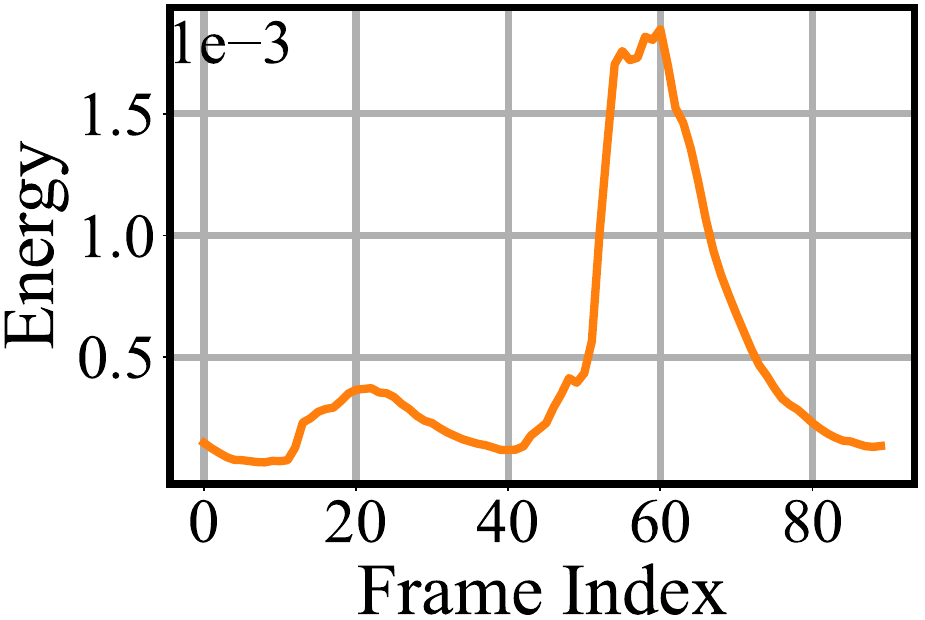}}
\caption{The Energy curve}
\label{fig1:1}
\end{subfigure}
\begin{subfigure}[t]{0.48\columnwidth}
\centerline{\includegraphics[width=0.99\columnwidth]{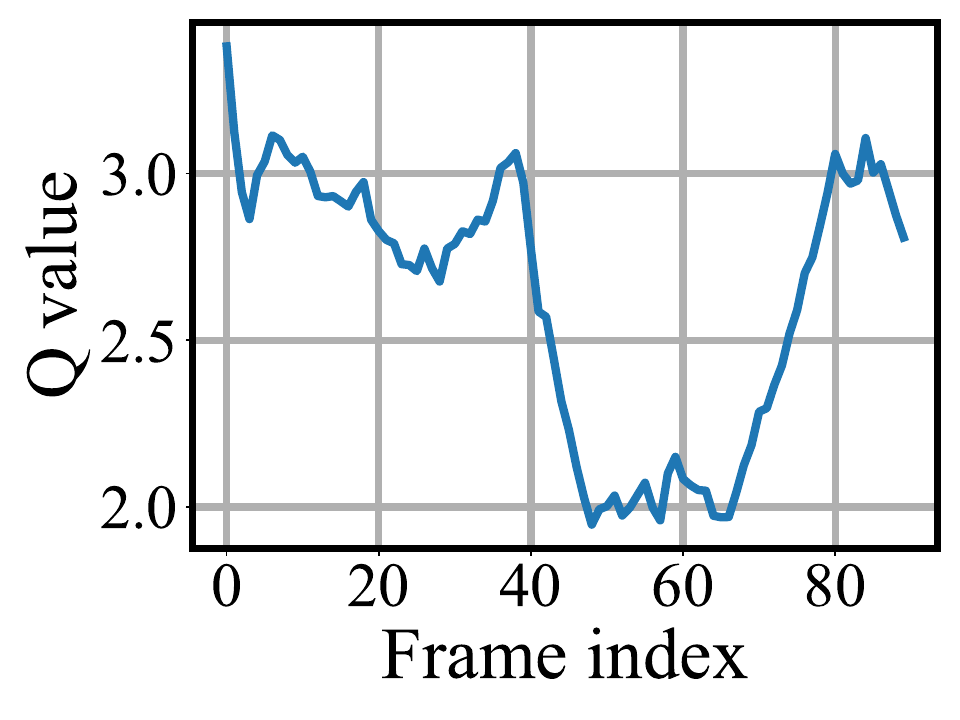}}
\caption{The Q value curve}
\label{fig1:2}
\end{subfigure}\caption{\textcolor{black}{Illustration of how (a) energy and (b) Q value change frame by frame for the Gabor filter with a center frequency of $3.46$ kHz (mid). Ada-FE infers one 1-second speech segment on the speaker ID task (VoxCeleb1).}}
\vspace{-0.5em}
\label{figQ2}
\end{figure}

\begin{figure}[!h]
% \vspace{-1.3em}
\centering
\begin{subfigure}[t]{0.49\columnwidth}
\centerline{\includegraphics[width=1.0\columnwidth]{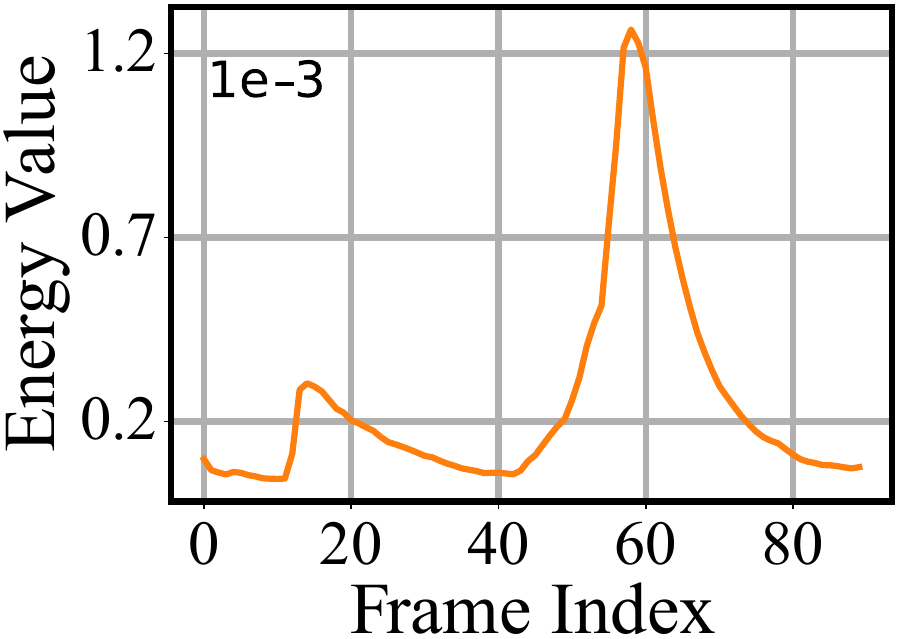}}
\caption{The Energy curve}
\label{fig1:1}
\end{subfigure}
\begin{subfigure}[t]{0.49\columnwidth}
\centerline{\includegraphics[width=0.99\columnwidth]{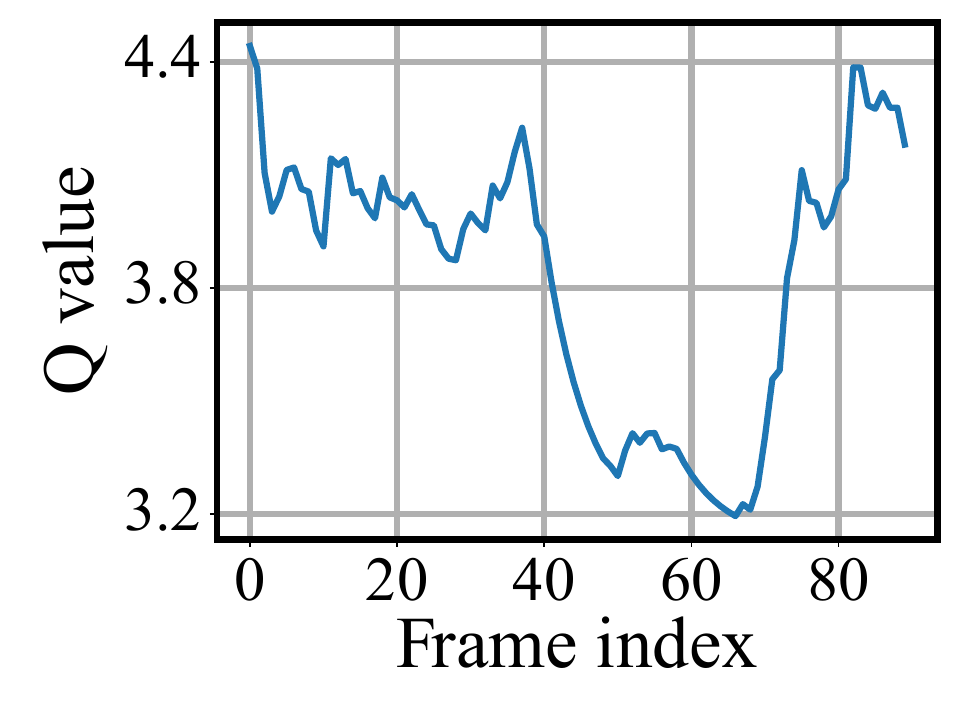}}
\caption{The Q value curve}
\label{fig1:2}
\end{subfigure}
\caption{\textcolor{black}{Illustration of how (a) energy and (b) Q value change frame by frame for the Gabor filter with a center frequency of $5.81$ kHz (high). Ada-FE infers one 1-second speech segment on the speaker ID task (VoxCeleb1).}}
\vspace{-0.5em}
\label{figQ21}
\end{figure}

\textcolor{black}{In Figures~\ref{figQ1}-\ref{figQ21}, we illustrate three examples to visualize how the value of Q-factor changes as the energy frame by frame at inference time, where our Ada-FE (with EfficientNet-B0 back-end classifier) infers 1-second audio segments for speaker ID (VoxCeleb1). The Q value and energy curves displayed in Figures~\ref{figQ1}-\ref{figQ21} correspond to the adaptive Gabor filters with center frequencies of $1.12$ kHz (low), $3.46$ kHz (mid), and $5.81$ kHz (high), respectively. It can be clearly observed that, overall, the Q-factor curve shows an opposite trend to the energy curve across different frequency channels. Given a frame with a low energy value, we expect Ada-FE to dynamically produce a corresponding high Q-factor value, leading to a high gain (sensitivity) and a low bandwidth (selectivity), and conversely, expect a low Q value, which is consistent with those shown in Figures~\ref{figQ1}-\ref{figQ21}. These observations about the relationship between energy and Q values over time illustrate how the proposed model adaptively responds to time-varying signal characteristics.}

\begin{figure}[!htbp]
% \vspace{-1.3em}
\centering
\captionsetup[sub]{font=normalsize}
\begin{subfigure}[t]{\columnwidth}
    % \centerline{\includegraphics[width=0.95\columnwidth]{AFC2v3.pdf}}
    \centerline{\includegraphics[width=0.95\columnwidth]{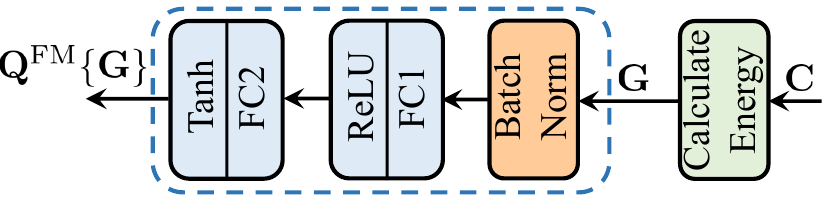}}
\caption{}
% \caption{AFC with the energy of each subband or channel as the input.}
\vspace{1.5em}
\label{fig1:1}
\end{subfigure}
\begin{subfigure}[t]{\columnwidth}
% \centerline{\includegraphics[width=0.95\columnwidth]{AFCSv4.pdf}}
\centerline{\includegraphics[width=0.95\columnwidth]{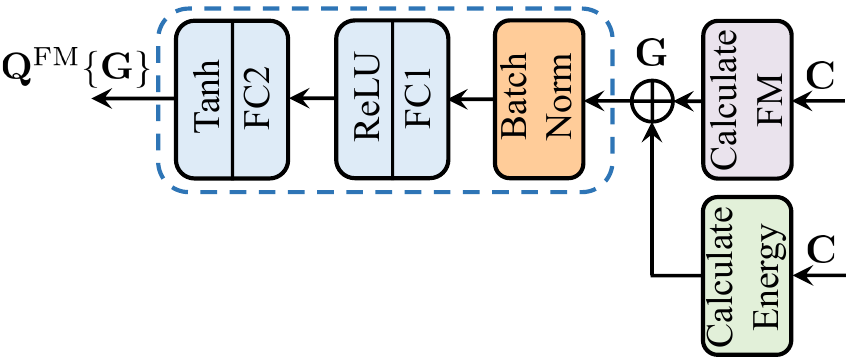}}
% \caption{AFC with the concatenation of the energy and FM as the input.}
\caption{}
\label{fig1:2}
\end{subfigure}
\caption{\textcolor{black}{The illustrations of the AFC with (a) the energy of each subband (AFC-Energy) and (b) the concatenation of the energy and FM as the input (AFC-Energy+FM), respectively. $\bigoplus$ represents the concatenation operation.}}
% \vspace{-1.5em}
\label{figafc}
\end{figure}

\input{Ablation}
% \subsection{Ablation Studies}
\subsection{\textcolor{black}{\textcolor{black}{Analyses of} Simplified Ada-FE}}\label{sec:5.2}

% \textcolor{black}{In this section, we perform comprehensive experiments to evaluate our Ada-FE and simplified Ada-FE (Ada-FE-S) across six audio and speech \textcolor{blue}{datasets}, with the EfficientNet-B0 as the back-end. For Ada-FE-S, we explore different design choices for the input to the adaptive feedback controller (AFC). We refer to the Ada-FE-S with the AFC that takes the original FM input (shown in Figure \ref{fig5}) as the Ada-FE-S-F. As shown in Figures \ref{figafc}(a) and (b), we also attempt to employ the energy of each subband and the concatenation of the energy and FM as the input to AFC, which are referred as Ada-FE-E and Ada-FE-EF, respectively.} 

In this section, we comprehensively evaluate the simplified Ada-FE (Ada-FE-S) across sound event classification, music genre classification, speech emotion classification, and keyword spotting tasks, with EfficientNet-B0 as the back-end. Built upon Ada-FE-S, we probe different design choices for the input to the neural adaptive feedback controller (AFC). We refer to the Ada-FE-S with the original AFC that takes the FM input (shown in Figure~\ref{fig5}) as the \textcolor{black}{Ada-FE-S-FM}. \textcolor{black}{As shown in Figure~\ref{figafc}\,(a) and (b), we also attempt to employ the energy of each subband and the concatenation of the energy and FM as the input to the AFC, which are referred to as \textcolor{black}{Ada-FE-S-Eg and Ada-FE-S-EgFM}, respectively.}

\textcolor{black}{The evaluation results are given in Table~\ref{ablation}, where Top-1 and Top-5 accuracy are presented. Top-1 and Top-5 accuracy respectively measure the proportion of times the top prediction matches the true label and the proportion of times true label is among the top five predictions. We can observe that Ada-FE-S-FM, proposed simplified (S) model with Frequency Modulation (FM) features as input, achieves comparable or better performance than the original Ada-FE on all the tasks, demonstrating that the Q-factor adaptive learning in Ada-FE benefits from removing the hand-crafted module. Among Ada-FE-S-FM (with FM input), Ada-FE-S-Eg (with Energy input, Figure~\ref{figafc}\,(a)), and Ada-FE-S-EgFM (with Energy and FM inputs, Figure~\ref{figafc}\,(b)), overall, Ada-FE-S-FM and Ada-FE-S-EgFM are superior to Ada-FE-S-Eg. We also can observe that the fusion of energy and FM features is not consistently beneficial for Ada-FE-S. Ada-FE-S-EgFM outperforms Ada-FE-S-FM on CREMA-D, SPC-V1, and SPC-V2 benchmarks, while Ada-FE-S-FM achieves better performance on ESC-50, GTZAN, and FMA-Small benchmarks.}

% We can observe that Ada-FE-S-F achieves comparable or better performance than Ada-FE on all the tasks, demonstrating that the Q-factor adaptive learning in Ada-FE can benefit from removing the hand-crafted module. \textcolor{black}{Among Ada-FE-S-F, Ada-FE-S-E, and Ada-FE-S-FE, overall, Ada-FE-S-F and Ada-FE-S-EF are superior to Ada-FE-S-E and the fusion of energy and FM features is not consistently beneficial for Ada-FE-S.} \textcolor{black}{Ada-FE-S-EF outperforms Ada-FE-S-F on CREMA-D, SPC-V1, and SPC-V2 datasets, while Ada-FE-S-F achieves better performance on ESC-50, GTZAN, and FMA-Small datasets.}

% in Ada-FE can benefit from removing the hand-crafted LDA module.

\textcolor{black}{Further, an interesting question arises: ``May we only use a single adaptive Gabor filter layer for Ada-FE?''. To answer this question, we explore removing the fixed Gabor filters (purple box in Figure~\ref{fig2}\,(a)) to study the effect of this change to assess its efficacy. Similarly, we employ the FM, the energy, and the concatenation of the energy and FM as the input to AFC. The evaluation results are denoted in gray (Table~\ref{ablation}). The comparison results to Ada-FE-S confirm the role of the fixed Gabor filters. Again, it can be observed that the FM and the concatenation input of energy and FM provide better performance than the energy input.} %\textcolor{black}{In addition, we can find that the adaptive front-end with only a single adaptive Gabor filter layer consistently benefits from the fusion of FM and energy input on all datasets.}

\subsection{\textcolor{black}{Comparative Studies}}

In this section, we compare Ada-FE and \textcolor{black}{Ada-FE-S-FM} with recent TD-fbanks and LEAF. In Table~\ref{top1}, we report the Top-1 test accuracy of the models across \textcolor{black}{all} eight audio and speech datasets and two back-end classifier backbone networks (i.e., MobileNetV2-100 and EfficientNet-B0). In addition, we report the results (marked in gray) of several advanced state-of-the-art audio representation learning models based on pre-training. Ada-FE and \textcolor{black}{Ada-FE-S-FM} (as well as LEAF and TD-fbanks) only involve spectral decomposition and low-level feature extraction, with a small number of trainable parameters. Note that they are not directly comparable to pre-training models that exploit a large DNN to capture high-level features and are pre-trained on massive external data. AST-Scratch denotes the audio spectrogram Transformer (AST) model~\cite{gong21b_interspeech} trained from scratch without pre-training. AST-AS and AST-AS+IM denote the AST models pre-trained on AudioSet~\cite{audioset} and AudioSet+ImageNet, respectively. PANNs~\cite{kong2020panns} is the CNN14 pre-trained on AudioSet. Wav2Vec2-LS960~\cite{wav2vec2} denotes the Wav2Vec2 model pre-trained on LibriSpeech-960 corpus~\cite{librispeech}. AST, PANNs, and Wav2Vec2 involve \textcolor{black}{87.0M}, 79.7M, and \textcolor{black}{315.4M} parameters, respectively. The model size of the back-end classifier varies with the size (number of classes) of the classification head on the top. From IEMOCAP (four classes) to VoxCeleb1 ($1\,251$ classes), the sizes of MobileNetV2-100 and EfficientNet-B0 range from \textcolor{black}{2.23M} and \textcolor{black}{4.01M} to \textcolor{black}{3.83M} and \textcolor{black}{5.61M}, respectively. \textcolor{black}{In contrast, Ada-FE (1.51K parameters), LEAF (0.32K parameters), and TD-fbanks (32K parameters) have negligible parameters compared to the back-ends.}

\input{./Top1Acc}
\input{./Top5Acc}

\textcolor{black}{From Table~\ref{top1}, we observe that Ada-FE and \textcolor{black}{Ada-FE-S-FM} outperform LEAF and TD-fbanks, across all the tasks and two back-end classifiers (MobileNetV2-100 and EfficientNet-B0). In music genre classification (GTZAN), for instance, Ada-FE and \textcolor{black}{Ada-FE-S-FM} provide 17.1\% and 19.68\%, and 14.4\% and 18.71\% relative Top-1 accuracy improvements over LEAF on MobileNetV2-100 and EfficientNet-B0, respectively. In speech emotion recognition (IEMOCAP), Ada-FE and \textcolor{black}{Ada-FE-S-FM} showcase 13.51\% and 16.95\%, and 6.37\% and 8.71\% relative Top-1 accuracy gains over LEAF, on MobileNetV2-100 and EfficientNet-B0 respectively. Compared to the other speech and audio tasks, VoxCeleb1 speaker ID ($1\,251$ classes) is a more challenging task. In speaker ID, with 1-second segments for training, Ada-FE and \textcolor{black}{Ada-FE-S-FM} improve LEAF with relative Top-1 accuracy of 34.06\% and 30.89\%, and 5.64\% and 2.63\% on MobileNetV2-100 and EfficientNet-B0, respectively. In addition, Ada-FE and \textcolor{black}{Ada-FE-S-FM} (with MobileNet-V2-100 and EfficientNet-B0 back-ends) exhibit better performance than the AST-Scratch (without pre-training), with significantly lower parameter overheads.}

% and 14.4\%, and 3.18\% and 9.19\% relative Top-1 accuracy improvements over LEAF on MobileNetV2-100 and EfficientNet-B0, on GTZAN and FMA-Small datasets respectively. Compared to the other speech and audio tasks, VoxCeleb1 speaker ID ($1\,251$ classes) is a more challenging task. In the speaker ID, with 1-second segments for training, Ada-FE improves LEAF with relative Top-1 accuracies of 34.06\% and 5.64\%, on MobileNetV2-100 and EfficientNet-B0, respectively. In addition, Ada-FE and Ada-FE-S-F (with MobileNet-V2-100 and EfficientNet-B0 back-ends) exhibit better performance than the AST-Scratch (without pre-training), with significantly lower parameter overheads.

% In music genre classification, for instance, Ada-FE provides 17.1\% and 14.4\%, and 3.18\% and 9.19\% relative Top-1 accuracy improvements over LEAF on MobileNetV2-100 and EfficientNet-B0, on GTZAN and FMA-Small datasets respectively. Compared to the other speech and audio tasks, VoxCeleb1 speaker ID ($1\,251$ classes) is a more challenging task. In the speaker ID, with 1-second segments for training, Ada-FE improves LEAF with relative Top-1 accuracies of 34.06\% and 5.64\%, on MobileNetV2-100 and EfficientNet-B0, respectively. In addition, Ada-FE and Ada-FE-S-F (with MobileNet-V2-100 and EfficientNet-B0 back-ends) exhibit better performance than the AST-Scratch (without pre-training), with significantly lower parameter overheads.

\textcolor{black}{Table~\ref{top5} reports the Top-5 accuracy of Ada-FE, \textcolor{black}{Ada-FE-S-FM}, and LEAF across seven datasets (except for IEMOCAP with four classes), on two back-ends. We can observe a performance trend similar to that shown in Table~\ref{top1}. In sound event classification (ESC-50) and speech emotion recognition (CREMA-D), Ada-FE attains 15.74\% and 9.51\%, 0.72\% and 0.89\% relative Top-5 accuracy improvements on MobileNetV2-100 and EfficientNet-B0, respectively. For VoxCeleb1 speaker ID, with 1-second training segments, Ada-FE improves LEAF with relative Top-5 accuracy of 20.08\% and 2.06\%, on the two back-ends respectively.}

% \textcolor{black}{Table I reports the Top-1 and Top-5 test accuracies (\%) of Ada-FE and LEAF, and Top-1 results of TD-fbank. It can be observed that Ada-FE demonstrates the best performance on two tasks. In the keyword spotting task, Ada-FE provides a relative Top-1 accuracy increase of 1.04\% over LEAF. Compared to keyword spotting (35 classes), VoxCeleb1 speaker Id. (1251 classes) is a more challenging task. In speaker Id. with 1-second segments for training, Ada-FE improves LEAF with relative Top-1 and Top-5 accuracies of 5.44\% and 2.36\%, respectively.}

% \vspace{-0.9em}

\begin{figure}[!t]
% \vspace{-0.7em}
  \centering
  \includegraphics[width=0.85\linewidth]{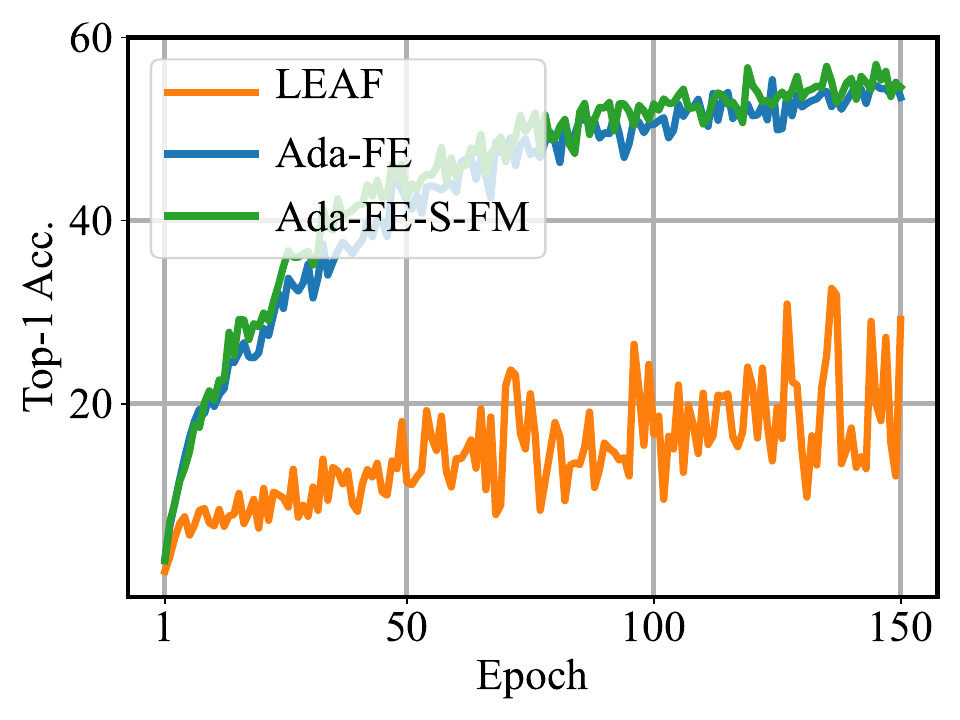}
  % \vspace{-1.8em}
  \caption{Top-1 accuracy (\%) of the LEAF, Ada-FE, and \textcolor{black}{Ada-FE-S-FM} (with the MobileNetV2-100 back-end) on sound event classification task (ESC-50 dataset), over various training epochs. Averaged accuracy over five-fold validation sets is reported.}
  \label{fig:esc50mbnetv2}
  % \vspace{-1.5em}
\end{figure}

\begin{figure}[!htbp]
  \centering
  \includegraphics[width=0.85\linewidth]{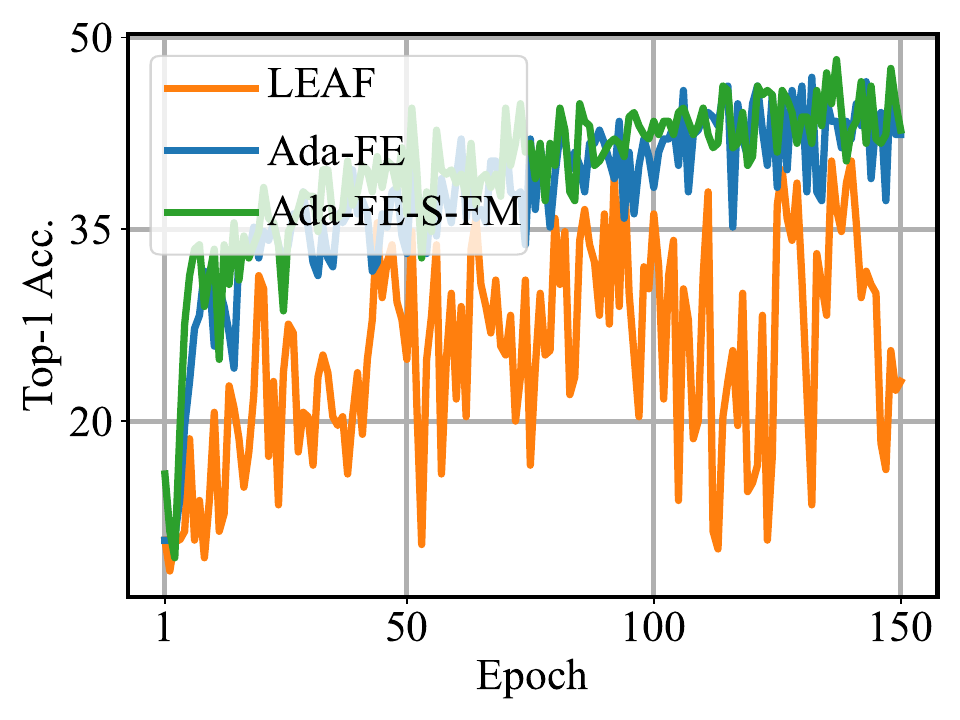}
  % \vspace{-0.8em}
  \caption{Top-1 accuracy (\%) of the LEAF, Ada-FE, and \textcolor{black}{Ada-FE-S-FM} (with the MobileNnetV2 back-end) on music genre classification task (GTZAN dataset), over various training epochs.}
  \label{fig:vox1mbnetv2}
  % \vspace{-0.9em}
\end{figure}

\begin{figure}[!h]
  \centering
  \includegraphics[width=0.85\linewidth]{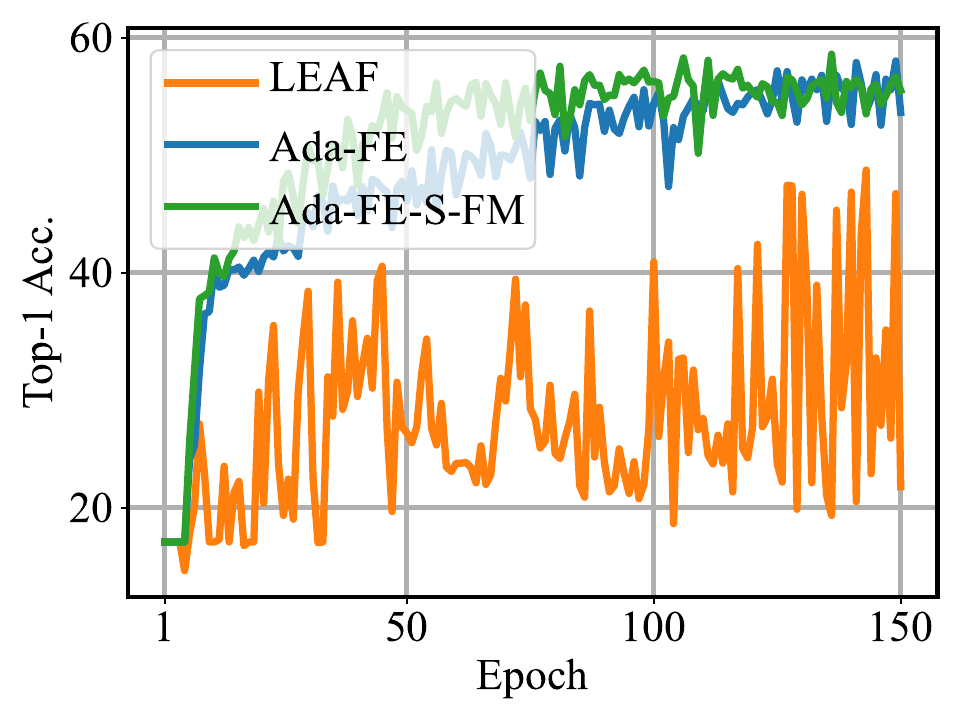}
  % \vspace{-0.8em}
  \caption{Top-1 accuracy (\%) of the LEAF, Ada-FE, and \textcolor{black}{Ada-FE-S-FM} (with the EfficientNet-B0 back-end) on speech emotion recognition (CREMA-D dataset), over various training epochs.}
  \label{fig:cremad}
  \vspace{-1.5em}
\end{figure}

\begin{figure}[!hpbt]
  \centering
  \includegraphics[width=0.85\linewidth]{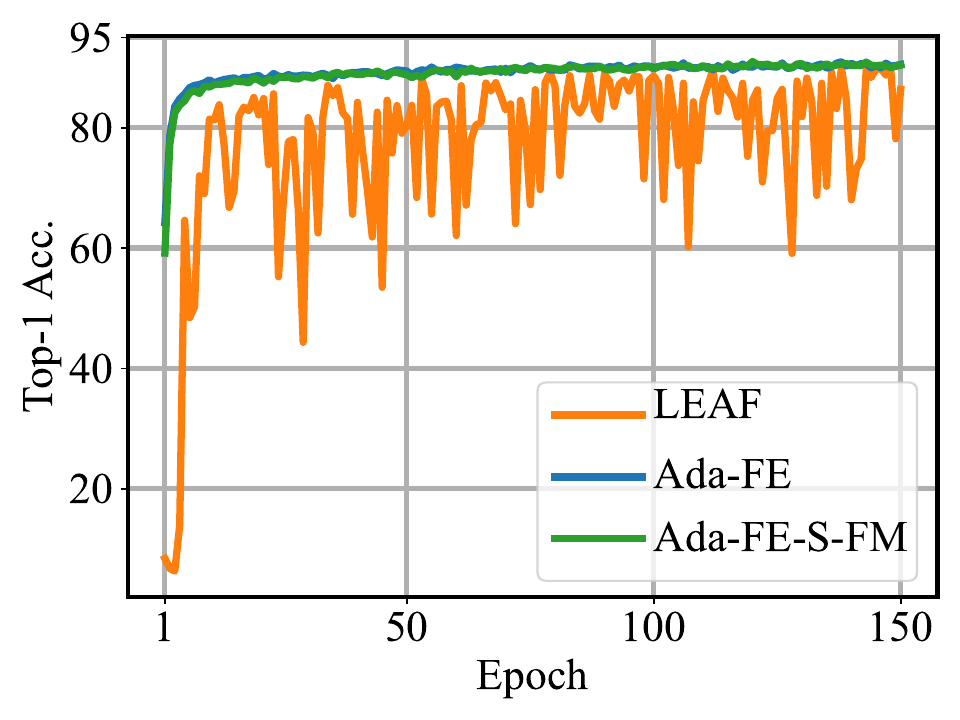}
  % \vspace{-0.8em}
  \caption{Top-1 accuracy (\%) of the LEAF, Ada-FE, and \textcolor{black}{Ada-FE-S-FM} (with the EfficientNet-B0 back-end) on keyword spotting task with the SPC-V2 dataset, over various training epochs.}
  \label{fig:scv2}
  \vspace{-0.9em}
\end{figure}

\textcolor{black}{Figures~\ref{fig:esc50mbnetv2}-\ref{fig:scv2} display the learning curves of our Ada-FE and \textcolor{black}{Ada-FE-S-FM} compared with that of the LEAF, on keyword spotting (SPC-V2), speaker ID (VoxCeleb1), speech emotion recognition (CREMA-D), and keyword spotting (SPC-V2), respectively. It can be easily observed that our Ada-FE and \textcolor{black}{Ada-FE-S-FM} demonstrate a faster increase in recognition accuracy, across MobileNetV2-100 and EfficientNet-B0 back-ends. As shown in Figure~\ref{fig:scv2}, for instance, Ada-FE and \textcolor{black}{Ada-FE-S-FM} with EfficientNet-B0 attain over 56\% accuracy with one-epoch training and converges well after 10-epoch training. The LEAF shows less than 10\% accuracy with one-epoch training. We also can find that the best performance attained by Ada-FE and \textcolor{black}{Ada-FE-S-FM} is better than that attained by LEAF.} 

Beyond the performance superiority, and perhaps more importantly, we find that our Ada-FE and \textcolor{black}{Ada-FE-S-FM} show much better stability over unseen test samples when compared with LEAF. This can be highly advantageous in a data-driven learning task. In real-world applications, for instance, speech signals captured by microphones are inevitably shaped by transfer channels between speakers and microphones. Transfer channels vary with the distance between speakers and microphones, as well as the acoustic scenarios (such as car space, church, hall, and outdoor). The learning curves of LEAF suggest that LEAF is quite sensitive to varying acoustic conditions. In contrast, the stability or robustness exhibited by the Ada-FE is significantly better and bodes well for practical applications.

\section{Conclusion}\label{sec6}

\textcolor{black}{\textcolor{black}{In this paper, we investigate the Ada-FE across two back-end classifiers and eight audio and speech classification benchmarks (including sound event classification, emotion recognition, keyword spotting, speaker identification, and music genre classification tasks).} Ada-FE consists of one fixed Gabor filter layer and an adaptive Gabor filter layer. The adaptive filters are dynamically tuned by hand-crafted level-dependent adaption and neural adaptive feedback controller. We further simplify the adaptive control process to enable the adaptive filters completely adjusted by the neural feedback controller.
% Furthermore, we study the adaptive front-end with only one adaptive Gabor filter layer.
}

Comprehensive experimental results confirm the performance superiority of Ada-FE over the recent state-of-the-art LEAF and suggest that using an adaptive front-end offers advantages over using a fixed front-end. Additionally, The simplified Ada-FE (\textcolor{black}{Ada-FE-S-FM}) provides highly competitive or better performance than Ada-FE, demonstrating hand-crafted LDA is quite essential. Beyond the performance superiority, Ada-FE and \textcolor{black}{Ada-FE-S-FM} show significantly better stability than the baseline on the test set, over various epochs. This paper demonstrates the advantages of employing feedback control in the front-end to respond to varying conditions by suitably altering filter characteristics, opening up new avenues for future research in adaptive speech and audio representations. In future work, we will further incorporate the active process into audio representation learning, such as pre-training models. 
\ifCLASSOPTIONcaptionsoff
  \newpage
\fi



\bibliographystyle{IEEEtran}
% \bibliography{IEEEabrv,../bib/paper}
% \scriptsize
% \newpage
\bibliography{IEEEabrv,./myreference}
\end{document}

%% file: dataset.tex
\begin{table*}[!htbp]
\centering
    % \scriptsize
    % \footnotesize
    \small
    \def\arraystretch{1.08}
    \setlength{\tabcolsep}{3.9pt}
    \setlength{\abovetopsep}{0pt}
    \setlength\belowbottomsep{0pt} 
    \setlength\aboverulesep{0pt} 
    \setlength\belowrulesep{0pt}
\caption{The details of eight datasets used for performance evaluation, including the number of classes, average duration, and dataset splits.
% \textcolor{blue}{Note, for TD-filterbank, we list the best Top-1 results from \cite{leaf}. For LEAF, the results are obtained with the official code, where we attain a lower keyword spotting and a higher speaker Id. Top-1 Acc. than that in the original paper.}
}
% \vspace*{-0.5em}
% \scalebox{1}{

\begin{tabular}{l|c|c|c|c|c|c|c|c}
% \toprule
\toprule[1.5pt]
% \hline
Task & Acoustic Scene & \multicolumn{2}{c|}{Music Genre} & \multicolumn{2}{c|}{Speech Emotion} & \multicolumn{2}{c|}{Keyword Spotting} & Speaker ID \\
\cline{2-9}
Dataset & ESC-50 & GTZAN & FMA-S & CREMA-D & IEMOCAP & SPC-V1 & SPC-V2 & VoxCeleb1 
\\
\toprule
\toprule
\# classes & 50 & 10 & 8 & 6 & 4 & 30 & 35 & $1\,251$ \\
Average duration & 5.0 s  & 30.0 s & 30.0 s & 2.5 s & 4.5 s & 1.0 s & 1.0 s & 8.2 s\\
\hline
\# train clips & \multirow{3}{*}{\makecell[c]{5 folds \\ $2\,000$}} & 443 & $6\,400$ & $5\,155$ & \multirow{3}{*}{\makecell[c]{5 folds \\ $5\,531$}} & $51\,088$ & $84\,843$ & $138\,316$ \\
\# validation clips &  & 197 & $800$ & $732$ &  & $6\,798$ & $9\,981$ & $6\,904$ \\
\# test clips &  & 290 & $800$ & $1\,551$ &  &  $6\,835$ & $11\,005$ & $8\,251$ \\

% \hline
\toprule[1.5pt]
\end{tabular}
\label{dataset}
% }
\end{table*}

%% file: Ablation.tex
\begin{table*}[!t]
\centering
    % \scriptsize
    % \footnotesize
    \small
    \def\arraystretch{1.3}
    \setlength{\tabcolsep}{2.3pt}
    \setlength{\abovetopsep}{0pt}
    \setlength\belowbottomsep{0pt} 
    \setlength\aboverulesep{0pt} 
    \setlength\belowrulesep{0pt}
\caption{\textcolor{black}{The Ada-FE involves a fixed Gabor filter layer and an adaptive Gabor filter layer that is dynamically tuned via the LDA and AFC modules. We start from the Ada-FE (Figure~\ref{fig2}\,(a)) and remove its features moving towards a front-end that only involves one adaptive Gabor filter layer controlled by the AFC: 1) removing LDA module (Ada-FE-S); 2) removing the fixed Gabor filters (shown in gray); 3) replacing FM feature (input for original AFC) with energy feature; 4) replacing FM feature with the concatenation of FM and energy. \textcolor{black}{We report Top-1 and Top-5 accuracy and indicate the best results on each dataset in boldface.} \textcolor{black}{IEMOCAP with four classes is not included.} \textcolor{black}{Here w/o denotes `without'.}} }
% \caption{\textcolor{red}{Disentangling Ada-FE. The Ada-FE involves a fixed Gabor filter layer followed by spatial differentiation, and an adaptive Gabor filter layer that dynamically tunes Q values via LDA and AFC modules. We start from the Ada-FE and remove its features moving towards a front-end that only involves one adaptive Gabor filter layer controlled by AFC (shown in gray): (1) removing LDA module; (2) removing fixed Gabor filters; (3) replacing FM feature (input for original AFC) with energy feature; (4) replacing FM feature with FM concatenated with energy. The Top-1 and Top-5 accuracy are reported, and the best accuracy values for each dataset are indicated in boldface numbers.}
% \textcolor{blue}{Note, for TD-filterbank, we list the best Top-1 results from \cite{leaf}. For LEAF, the results are obtained with the official code, where we attain a lower keyword spotting and a higher speaker Id. Top-1 Acc. than that in the original paper.}

% \vspace*{-0.5em}
% \scalebox{1}{
\begin{tabular}{l|cc|cc|cc|cc|cc|cc}
\toprule[1.5pt]
% \hline
% Task & Acoustic Scene & \multicolumn{2}{c|}{Music Genre} & \multicolumn{2}{c|}{Speech Emotion} & \multicolumn{2}{c|}{keyword Spotting} & Speaker Id. \\
% \hline
\multirow{2}{*}{Model} & \multicolumn{2}{c|}{ESC-50} & \multicolumn{2}{c|}{GTZAN} &  \multicolumn{2}{c|}{FMA-Small} & \multicolumn{2}{c|}{CREMA-D} & \multicolumn{2}{c|}{SPC-V1} & \multicolumn{2}{c}{SPC-V2} \\
% \hline
\cline{2-13}
& Top-1 & Top-5 & Top-1 & Top-5 & Top-1 & Top-5 & Top-1 & Top-5 & Top-1 & Top-5 & Top-1 & Top-5 \\
% \hline
% Segment & -- & -- & -- & -- & -- & -- & -- & 1 s & 3 s \\
\hline
\hline
Ada-FE (Fixed Gabor filters+LDA+AFC-FM) & \textcolor{black}{64.80} & \textcolor{black}{88.00} & \textcolor{black}{54.83} & \textcolor{black}{95.52} & \textcolor{black}{46.00} & \textcolor{black}{93.00} & \textcolor{black}{58.03} & \textcolor{black}{98.97} & 91.29 & 97.83 & \textcolor{black}{90.79} & \textcolor{black}{97.89} \\
\hline

\,\,\, \textcolor{black}{w/o Adaptation (Not Learnable)} & \textcolor{black}{49.75} & \textcolor{black}{79.25} & \textcolor{black}{45.52} & \textcolor{black}{89.66} & \textcolor{black}{38.88} & \textcolor{black}{91.00} & \textcolor{black}{48.55} & \textcolor{black}{97.68} & \textcolor{black}{87.84} & \textcolor{black}{96.98} & \textcolor{black}{87.23} & \textcolor{black}{97.30} \\
\hline
\textcolor{black}{Ada-FE-S-FM} (w/o LDA, AFC-FM) & \textbf{\textcolor{black}{65.30}}  & \textcolor{black}{88.05} & \textbf{\textcolor{black}{56.90}}  & \textbf{\textcolor{black}{96.21}} & \textbf{\textcolor{black}{47.38}} & \textbf{\textcolor{black}{93.38}} & \textcolor{black}{58.61} & \textbf{\textcolor{black}{99.03}} & 91.14  &  97.66  & \textcolor{black}{90.91} & \textcolor{black}{97.83} \\

\textcolor{black}{Ada-FE-S-Eg} (w/o LDA, AFC-Energy) & \textcolor{black}{64.55}  & \textbf{\textcolor{black}{88.40}} & \textcolor{black}{50.00}  & \textcolor{black}{92.41} & \textcolor{black}{43.75} & \textcolor{black}{90.63} & \textcolor{black}{58.26} & \textcolor{black}{98.52} & 91.49 & 97.98  & \textcolor{black}{91.31} & \textcolor{black}{97.92} \\

\textcolor{black}{Ada-FE-S-EgFM} (w/o LDA, AFC-Energy+FM) & \textcolor{black}{64.85}  & \textcolor{black}{88.25} & \textcolor{black}{54.83}  & \textcolor{black}{95.17} & \textcolor{black}{47.13} & \textbf{\textcolor{black}{93.38}} & \textbf{\textcolor{black}{58.99}} & \textbf{\textcolor{black}{99.03}} & \textbf{92.64} & \textbf{98.14} & \textbf{\textcolor{black}{92.09}} & \textbf{\textcolor{black}{98.15}} \\

\toprule

\rowcolor{gray!10} \,\,\,\,\, w/o Fixed Gabor filters (AFC-FM)    
& \textcolor{black}{59.60} & \textcolor{black}{86.20} & \textcolor{black}{50.00} & \textcolor{black}{91.38} & \textcolor{black}{40.00} & \textcolor{black}{91.75} & \textcolor{black}{53.84} & \textcolor{black}{98.32} & 82.52 & 95.49 & \textcolor{black}{80.12} & \textcolor{black}{94.81} \\

\rowcolor{gray!10} \,\,\,\,\, w/o Fixed Gabor filters (AFC-Energy) & \textcolor{black}{57.95} & \textcolor{black}{85.65} & \textcolor{black}{47.93} & \textcolor{black}{89.97} & \textcolor{black}{35.88} & \textcolor{black}{90.75} & \textcolor{black}{47.84} & \textcolor{black}{98.26} & 73.41 & 91.77 & \textcolor{black}{72.21} & \textcolor{black}{91.30} \\

\rowcolor{gray!10} \,\,\,\,\, w/o Fixed Gabor filters (AFC-Energy+FM) & \textcolor{black}{61.20} & \textcolor{black}{86.20} & \textcolor{black}{50.34} & \textcolor{black}{93.10} & 40.25 & 92.88 & 55.25 & 98.84 & 87.65 & 96.62 & \textcolor{black}{85.06} & \textcolor{black}{95.96} \\

\toprule[1.5pt]
% \hline
% \rowcolor{orange!20} \,\,\,\,\, Fixed Gabor filter+AFC-Energy+FM & -- & -- & 54.14 & 95.52 & 45.38 & 93.63 & 59.78 & 98.84 & 91.92 & 98.19 \\
\end{tabular}
\label{ablation}
% }
\end{table*}

%% file: Top1Acc.tex
\renewcommand{\thefootnote}{\fnsymbol{footnote}}

\begin{table*}[!t]
\vspace{1.0em}
\centering
% \centering
    % \scriptsize
    % \footnotesize
    \small
    \def\arraystretch{1.05}
    \setlength{\tabcolsep}{2.0pt}
    \setlength{\abovetopsep}{0pt}
    \setlength\belowbottomsep{0pt} 
    \setlength\aboverulesep{0pt} 
    \setlength\belowrulesep{0pt}
\caption{
\textcolor{black}{The \textbf{Top-1} accuracy of different models across eight datasets, on two back-end classifier netwroks. The best accuracy values for each dataset are indicated in boldface numbers. MNetV2 and ENet-B0 represent MobileNetV2-100 and EfficientNet-B0 back-end network models, respectively. \textcolor{black}{Here, we also list the results of several advanced audio representation methods based on pre-training (marked in gray).} $^{**}$ The results of AST-Scratch and AST-AS are adopted from Gong et al.~\cite{gong2022ssast}. $^\dagger$ The results of PANNs, AST-AS+IM, and Wav2Vec2-LS960 are adopted from Niizumi et al.~\cite{byol}. }
% \textcolor{blue}{Note, for TD-filterbank, we list the best Top-1 results from \cite{leaf}. For LEAF, the results are obtained with the official code, where we attain a lower keyword spotting and a higher speaker Id. Top-1 Acc. than that in the original paper.}
}
% \vspace*{-0.5em}
% \scalebox{1}{
% \centering

\begin{tabular}{l|c|c|c|c|c|c|c|c|c|c|c}
\toprule[1.5pt]
% \hline
% Task & Acoustic Scene & \multicolumn{2}{c|}{Music Genre} & \multicolumn{2}{c|}{Speech Emotion} & \multicolumn{2}{c|}{keyword Spotting} & Speaker Id. \\
% \hline
Method & Pre-train & Back-end & ESC-50 & GTZAN & FMA-S & CREMA-D & IEMOCAP & SPC-V1 & SPC-V2 & \multicolumn{2}{c}{VoxCeleb1} \\
% \hline
% Segment & -- & -- & -- & -- & -- & -- & -- & 1 s & 3 s \\
\hline
\hline
% TD-fbanks & \multirow{3}{*}{MNet-V2} & -- & -- & -- & -- & -- & -- & 87.70 & 26.0 (1s) & -- \\
LEAF & \multirow{3}{*}{\ding{56}} &  \multirow{3}{*}{MNetV2} & 
\textcolor{black}{41.20}  & \textcolor{black}{40.34} & \textcolor{black}{39.88}  & \textcolor{black}{49.65} & \textcolor{black}{49.68} & \textcolor{black}{73.38} & \textcolor{black}{82.35} & \textcolor{black}{25.54} (1s) & \textcolor{black}{38.78} (3s) \\

Ada-FE & & 
& \textcolor{black}{59.40}  & \textcolor{black}{47.24} & \textcolor{black}{41.25} & \textcolor{black}{57.51} & \textcolor{black}{56.39} & \textcolor{black}{\bf 84.45} & \textcolor{black}{84.63} & \textbf{\textcolor{black}{34.24}} (1s) & \textbf{\textcolor{black}{46.25}} (3s) \\

\textcolor{black}{Ada-FE-S-FM} & & 
& \textcolor{black}{\textbf{59.85}}  & \textbf{\textcolor{black}{48.28}} & \textbf{\textcolor{black}{44.50}} & \textbf{\textcolor{black}{58.16}} & \textbf{\textcolor{black}{58.05}} & \textcolor{black}{83.51} & \textbf{84.68} & \textcolor{black}{33.43} (1s) & \textcolor{black}{44.79} (3s) \\
\hline

TD-fbanks & \multirow{4}{*}{\ding{56}} & \multirow{4}{*}{ENet-B0} 
& -- & -- & -- & -- & -- & -- & 87.70 & 26.0 (1s) & -- \\
LEAF & & &
\textcolor{black}{51.20}  & \textcolor{black}{47.93}& 42.13  & \textcolor{black}{51.06} & \textcolor{black}{53.40} & 89.98 & 89.90 & 36.89 (1s) & 44.25 (3s) \\

Ada-FE & & &
64.80  & 54.83 & 46.00 & 58.03 & 56.80 & \textbf{91.29} & 90.79 & \textbf{38.97} (1s) & \textbf{48.69} (3s) \\

\textcolor{black}{Ada-FE-S-FM}   & & &
\textbf{65.30}  & \textbf{56.90} & \textbf{47.38} & \textbf{58.61} & \textbf{58.05} & 91.14 & \textbf{90.91} & \textcolor{black}{37.86} (1s) & \textcolor{black}{47.15} (3s) \\
\hline
\hline
AST-Scratch$^{**}$ & \ding{56}
& -- & 41.9 & -- & -- & -- & 51.9 & -- & 92.6 & \multicolumn{2}{c}{30.1} \\

\rowcolor{gray!20}
PANNs$^\dagger$ & 
& -- & 90.5 & 82.8 & -- & 50.9 & -- & -- & 51.7 & \multicolumn{2}{c}{8.4} \\

\rowcolor{gray!20}
AST-AS$^{**}$  &
& -- & 86.8 & -- & -- & -- & 51.9 & -- & 96.2 & \multicolumn{2}{c}{35.2} \\

\rowcolor{gray!20}AST-AS+IM$^\dagger$ & \multirow{-2}{*}{\ding{52}}
& -- & 93.9 & 86.1 & -- & 58.5 & -- & -- & 72.2 & \multicolumn{2}{c}{16.9} \\
\rowcolor{gray!20} 
Wav2Vec2-LS960$^\dagger$ &
& -- & 58.4 & 59.1 & -- & 67.2 & -- & -- & 96.6 & \multicolumn{2}{c}{41.5} \\
% \hline
\toprule[1.5pt]
\end{tabular}
\label{top1}
% }
\end{table*}

%% file: Top5Acc.tex
\begin{table*}[!htbp]
\vspace{0.8em}
\centering
% \centering
    % \scriptsize
    % \footnotesize
    \small
    \def\arraystretch{1.2}
    \setlength{\tabcolsep}{2.5pt}
    \setlength{\abovetopsep}{0pt}
    \setlength\belowbottomsep{0pt} 
    \setlength\aboverulesep{0pt} 
    \setlength\belowrulesep{0pt}
\caption{\textcolor{black}{The \textbf{Top-5} accuracy of Ada-FE, Ada-FE-S-F, and LEAF across seven datasets (IEMOCAP with four classes is not included), on two back-end classifiers. The best accuracy scores for each dataset are indicated in boldface numbers. MNetV2 and ENet-B0 represent MobileNetV2-100 and EfficientNet-B0 back-end classifiers, respectively.}
% \textcolor{blue}{Note, for TD-filterbank, we list the best Top-1 results from \cite{leaf}. For LEAF, the results are obtained with the official code, where we attain a lower keyword spotting and a higher speaker Id. Top-1 Acc. than that in the original paper.}
}
% \vspace*{-0.5em}
% \scalebox{1}{
% \centering
\begin{tabular}{l|c|c|c|c|c|c|c|c|c}
\toprule[1.5pt]
% \hline
% Task & Acoustic Scene & \multicolumn{2}{c|}{Music Genre} & \multicolumn{2}{c|}{Speech Emotion} & \multicolumn{2}{c|}{keyword Spotting} & Speaker Id. \\
% \hline
Method & Back-end & ESC-50 & GTZAN & FMA-S & CREMA-D & SPC-V1 & SPC-V2 & \multicolumn{2}{c}{VoxCeleb1} \\
% \hline
% Segment & -- & -- & -- & -- & -- & -- & -- & 1 s & 3 s \\
\hline
\hline
% TD-fbanks & \multirow{3}{*}{MNet-V2} & -- & -- & -- & -- & -- & -- & 87.70 & 26.0 (1s) & -- \\
LEAF     &  \multirow{3}{*}{MNetV2} & 
\textcolor{black}{74.95}  & \textcolor{black}{89.66}& \textcolor{black}{91.88}  & \textcolor{black}{98.19} & \textcolor{black}{93.25} & \textcolor{black}{95.47} & \textcolor{black}{41.07} (1s) & \textcolor{black}{59.51} (3s) \\

Ada-FE    & & 
\textcolor{black}{84.00}  & \textcolor{black}{93.45} & \textcolor{black}{92.63} & \textcolor{black}{98.68} & \textbf{95.90} & \textbf{\textcolor{black}{95.92}} & \textbf{\textcolor{black}{51.39}} (1s) & \textbf{\textcolor{black}{66.28}} (3s) \\

\textcolor{black}{Ada-FE-S-FM}  & & 
\textbf{\textcolor{black}{85.80}}  & \textbf{\textcolor{black}{93.90}} & \textbf{\textcolor{black}{93.00}} & \textbf{\textcolor{black}{98.84}} & \textcolor{black}{95.84} & \textcolor{black}{95.78} & \textcolor{black}{50.53} (1s) & \textcolor{black}{65.35} (3s) \\
% \textbf{\textcolor{blue}{83.97}} IEMOCAP
\hline
% TD-fbanks & \multirow{3}{*}{ENet-B0} & -- & -- & -- & -- & -- & -- & 87.70 & 26.0 (1s) & -- \\

LEAF     & \multirow{3}{*}{ENet-B0} & 
\textcolor{black}{82.00}  & \textcolor{black}{91.72}& \textcolor{black}{92.75}  & \textcolor{black}{98.13} & \textcolor{black}{97.10} & \textcolor{black}{97.54} & \textcolor{black}{54.76} (1s) & \textcolor{black}{64.58} (3s) \\
% \textcolor{blue}{83.62} IEMOCAP

Ada-FE    & & 
\textcolor{black}{88.00} & \textcolor{black}{95.52} & \textcolor{black}{93.00} & \textcolor{black}{98.97} & \textbf{\textcolor{black}{97.83}} & \textbf{\textcolor{black}{97.89}} & \textbf{\textcolor{black}{55.97}} (1s) & \textbf{\textcolor{black}{69.14}} (3s) \\

\textcolor{black}{Ada-FE-S-FM}    & & 
\textbf{\textcolor{black}{88.05}}  & \textbf{\textcolor{black}{96.21}} & \textbf{\textcolor{black}{93.38}} & \textbf{\textcolor{black}{99.10}} & \textcolor{black}{97.66} & \textcolor{black}{97.87} & \textcolor{black}{55.37} (1s) & \textcolor{black}{67.73} (3s) \\

% \textbf{\textcolor{blue}{84.46}} IEMOCAP
% \hline
% \hline
% \rowcolor{gray!20} AST-Scratch 
% & -- & 41.9 & -- & -- & -- & 51.9 & -- & 92.6 & \multicolumn{2}{c}{30.1} \\
% \rowcolor{gray!20} AST-AS  
% & -- & 86.8 & -- & -- & -- & 51.9 & -- & 96.2 & \multicolumn{2}{c}{35.2} \\ 
% \rowcolor{gray!20} AST-AS+IM
% & -- & 93.9 & 86.1 & -- & 58.5 & -- & -- & 72.2 & \multicolumn{2}{c}{16.9} \\ 
% \rowcolor{gray!20} Wav2Vec2-LS960
% & -- & 67.3 & 63.3 & -- & 56.9 & -- & -- & 86.0 & \multicolumn{2}{c}{32.5} \\
% \hline
\toprule[1.5pt]
\end{tabular}
\label{top5}
% }
\end{table*}